\documentclass[12pt,a4paper]{article}

\usepackage{epsf}
\usepackage{amsmath}
\usepackage{bbm}
\usepackage{rotate}
\usepackage{rotating}
\usepackage[hypertex]{hyperref}
\renewcommand{\baselinestretch}{1.5}
\makeatletter \@addtoreset{equation}{section} \makeatother

\newcommand{\be}{\begin{equation}}
\newcommand{\ee}{\end{equation}}
\newcommand{\bea}{\begin{eqnarray}}
\newcommand{\eea}{\end{eqnarray}}
\newcommand{\cN}{\mathcal{N}}
\newcommand{\cL}{\mathcal{L}}
\newcommand{\cZ}{\mathcal{Z}}
\newcommand{\pV}{\mathbbm{V}}

\newcommand{\cD}{\mathcal{D}}
\newcommand{\cR}{\mathcal{R}}
\newcommand{\cW}{\mathcal{W}}
\newcommand{\cM}{\mathcal{M}}
\newcommand{\cH}{\mathcal{H}}
\newcommand{\C}{\mathbbm{C}}

\newcommand{\Rom}{\mathbbm{R}}
\newcommand{\Ib}{\bar{I}}
\newcommand{\Jb}{\bar{J}}
\newcommand{\zp}{z_{+}}
\newcommand{\zbp}{\bar{z}_{+}}
\newcommand{\unit}{{\mathbbm{1}}}
\newcommand{\imag}{i}
\newcommand{\rmd}{\mbox{\rm{d}}}
\newcommand{\e}{\mbox{\rm{e}}}
\newcommand{\ub}{\bar{u}}
\newcommand{\vb}{\bar{v}}
\newcommand{\jb}{\bar{\jmath} }
\newcommand{\ib}{\bar{\imath} }
\newcommand{\kb}{\bar{k}}
\newcommand{\lb}{\bar{l}}
\newcommand{\zb}{\bar{z}}

\newcommand{\tI}{\tilde{I}}
\newcommand{\tJ}{\tilde{J}}
\newcommand{\tK}{\tilde{K}}

\newcommand{\kh}{\hat{k}}
\newcommand{\h}{\tilde{h}}
\newcommand{\mscr}[1]{\mbox{\scriptsize #1}}
\newcommand{\ft}[2]{{\textstyle\frac{#1}{#2}}}
\begin{document}
\begin{titlepage}
\begin{center}

\hfill hep-th/0310173\\
\hfill FSU-TPI-08/03

\vskip 1cm {\large \bf  Effective Supergravity Actions for  Flop Transitions
}\footnote{Work supported by the `Schwerpunktprogramm Stringtheorie' of the DFG.}

\vskip .5in

{\bf Laur J\"arv, Thomas Mohaupt and Frank Saueressig }  \\

{\em Institute of Theoretical Physics,
Friedrich-Schiller-University 
Jena, \\ 
 Max-Wien-Platz 1, D-07743 Jena, Germany}\\
{\tt L.Jaerv, T.Mohaupt, F.Saueressig@tpi.uni-jena.de}

\vskip 0.5cm

\end{center}
\vskip 1.5cm

\begin{center} {\bf ABSTRACT } \end{center}

\noindent 
We construct a family of five-dimensional gauged supergravity actions which
 describe flop transitions
of M-theory compactified on Calabi-Yau threefolds. While the vector multiplet
sector can be treated exactly, we use the Wolf spaces
$X(1+N) = \frac{U(1+N,2)}{U(1+N) \times U(2)}$ to model
the universal hypermultiplet together with  $N$ charged hypermultiplets
corresponding to winding states of the M2-brane. The metric,
the Killing vectors and the moment maps of these spaces are obtained
explicitly by using the superconformal quotient construction
of quaternion-K\"ahler manifolds. The inclusion of the extra hypermultiplets gives rise to a non-trivial scalar potential which is uniquely fixed by M-theory physics.

\end{titlepage}

\tableofcontents

\begin{section}{Introduction}

Supergravity actions provide a powerful tool for 
studying the low energy dynamics of string and M-theory 
compactified on special holonomy manifolds. In this case one usually has a moduli space of vacua, corresponding
to the deformations of the internal manifold $X$ and the background fields. For theories with
eight or less supercharges this moduli space
includes special points where $X$ becomes singular, leading to
a discontinuous or singular low energy effective action (LEEA).
However, within the full string
or M-theory these singularities are believed to be artifacts, which result
from ignoring some relevant modes of the theory, namely the winding states
of strings or branes around the cycles of $X$. Singularities of $X$ 
arise when such cycles are contracted to zero volume, which leads
to additional massless states. It was the crucial insight of
\cite{Str1} that the singularities occurring in the LEEA 
of type II strings compactified on a Calabi-Yau (CY) threefold with 
a conifold singularity can be interpreted as arising from illegitimately
integrating out such massless states. This has been generalized to many
other situations, including M-theory compactifications on CY
threefolds \cite{Witten1}. In some cases it is possible to resolve the
singularity of $X$ 
in two or more topologically different ways. This gives rise
to so-called topological phase transitions. Such transitions 
have been studied intensively in literature
\cite{GreFlop,WitD=2,Str2,Witten1}.\footnote{We refer to 
\cite{GreRev} for a review and more references.} They can 
be realized 
as parametric deformations of vacua, but also 
dynamically \cite{GMMS,GSS,MohProc,flop,us}. 

The usual LEEA only include those states which are generically massless, while the extra light modes occurring in a topological phase transition are left out. We refer to this description as the `Out-picture'. For the complete description of the low energy physics, however, one also needs to include the additional light modes. Following \cite{flop}, we will call these additional light modes `transition states'. The low energy description which explicitly includes the transition states will be referred to as the `In-picture'.

There are various reasons why it is important to know the In-picture description of topological phase transitions.  The compactification of type II string theory or  M-theory on smooth spaces gives rise to a massless
spectrum which only contains neutral states. However, in the vicinity
of special points one can get non-abelian gauge groups and charged
chiral matter, which makes such compactifications 
viable for particle physics
model building.  Since in these models all charged particles are 
transition states, it is clear that one needs the extended LEEA corresponding to the In-picture to describe their dynamics.
It has also been shown that in compactifications with background
flux the scalar potential has its minima at special points
in moduli space, where additional light states occur \cite{FluxComp,BerlinFlux}. 
Conversely, it has been noticed
in \cite{flop} that even in the absence of flux the potential
generated by the transition states has the effect
that the region in the vicinity of a topological phase transition is dynamically
preferred.
Finally, there is some evidence that the interplay between singularities and
background flux generates a small scale, which could help to solve
the gauge hierarchy  and the cosmological
constant problem \cite{Mayr,GKP,FMSW}.

Although
it is clear in principle that one should be able to ``integrate in''
the additional states, not much effort has been devoted towards working
out the corresponding LEEA explicitly. A systematic investigation
was started in  \cite{SU2} and continued in \cite{LMZ}, by deriving
the explicit LEEA which describe $SU(2)$ gauge symmetry enhancement
through string or brane winding states in five and four 
dimensions. For compactifications with
${\cal N}=4$ supersymmetry (16 supercharges) non-abelian gauge symmetry enhancement of the LEEA
has been considered in \cite{xx}.

The first step to obtain analogous results for flop transitions occurring in M-theory compactified on CY threefolds has been made in \cite{flop}. In this case the transition states are given by charged hypermultiplets which combine with the neutral hypermultiplets arising from the smooth CY compactification. Local supersymmetry requires that these fields parametrize a non-flat quaternion-K\"ahler manifold \cite{BW}. In \cite{flop} the difficulties in working with these rather complicated manifolds were avoided by taking the hypermultiplet manifold to be flat. This, however, is only compatible with {\it global} supersymmetry and does not give rise to a consistent supergravity description of the transition.

In this paper we construct In-picture LEEA for flop transitions which are $\cN = 2$ {\it locally} supersymmetric. The strategy  is to combine information about the transition states coming from M-theory with knowledge about the general $\cN = 2, D = 5$ gauged supergravity action \cite{VSG,GZ,gs,EGZ}.\footnote{Here ${\cal N}$ counts real
supercharges in multiples of 4. Thus ${\cal N}=2$ refers
to the smallest supersymmetry algebra in five dimensions.} 

As long as the CY threefold $X$ is smooth, the LEEA 
can be obtained by dimensional reduction \cite{CY_1}. Besides the
five-dimensional supergravity multiplet, it contains 
vector and hypermultiplets whose couplings are
determined by $X$. The LEEA is an ungauged supergravity action:
all fields are neutral, the gauge group is abelian, and there is no 
scalar potential. In a flop transition the K\"ahler moduli
are varied such that $X$ becomes singular through the contraction of
$N$ isolated holomorphic curves \cite{Witten1}. 
 The winding states of M2-branes
around these curves give rise to $N$ charged
hypermultiplets, which become massless at the transition locus.
These are the transition states that we want to integrate in.
Since they are charged, the resulting action
is a gauged supergravity action, which has a non-trivial
scalar potential.

The vector multiplet sector of the LEEA contains the K\"ahler moduli which control
the sizes of the $N$ holomorphic curves and, hence, the
phase transition. These parametrize a so-called
very special real manifold which is completely determined by
a cubic polynomial, the prepotential. In the Out-picture the
prepotential can be computed exactly and the threshold corrections arising from integrating out the 
transition states have  been derived in \cite{Witten1}.
As a result, we can determine the vector multiplet part of the In-picture
LEEA exactly. 

The situation is much more complicated in the hypermultiplet
sector, and this is the main point we have to address
in this paper. Local supersymmetry
requires that the hypermultiplet manifold is a quaternion-K\"ahler
manifold with non-trivial Ricci curvature \cite{BW}. The latter constraint
excludes hyper-K\"ahler and in particular flat manifolds.
The main difference 
between (generic) quaternion-K\"ahler manifolds and other geometries
familiar from supersymmetric theories, such as hyper-K\"ahler and
special K\"ahler manifolds, is that there are no simple, globally defined
holomorphic objects which encode the information one needs
to construct the LEEA. There is no K\"ahler potential
and in general a quaternion-K\"ahler manifold is not even
a complex manifold. Moreover, for the study of gaugings it would
be convenient to take the hypermultiplet manifold to be a 
direct product, with the neutral fields in one factor and the
charged fields in the other. But this is also not an option,
because the product of two (generic) quaternion-K\"ahler manifolds
is not quaternion-K\"ahler. 

Due to these complications, this type of geometry
is much less understood than the other geometries occurring in
supergravity. In particular, only very limited results
exist on how to explicitly compute the hypermultiplet metric
in string or M-theory. The best studied subsector is the 
universal hypermultiplet, which at tree level is described 
by the coset $\frac{U(1,2)}{U(1) \times U(2)}$, but  receives a non-trivial loop
correction \cite{AMTV}. The tree level result for the neutral hypermultiplets can be obtained through the c-map \cite{cmap,FS},
but only little is known about quantum corrections (see \cite{AspHM} for a review). 
Charged multiplets
have not been studied at all.

Therefore we take the approach
of using a toy model: to describe a flop transition we use 
a particular family of symmetric quaternion-K\"ahler spaces,
the non-compact versions of the unitary Wolf spaces
\be
\label{1.1}
X(1+N) = \frac{U(1+N,2)}{U(1+N) \times U(2)} \, ,
\ee
containing $N+1$ hypermultiplets. One of these hypermultiplets will be identified with  the
universal hypermultiplet while the $N$ other hypermultiplets 
will correspond to the transition states. Their charges are
determined by the geometry of the flop transition. The remaining neutral hypermultiplets
present in a generic LEEA will be ignored. This is a 
reasonable approximation, because these hypermultiplets
parametrize the complex structure of $X$, which is kept fixed
in a flop transition.  As we will show, this input suffices to uniquely determine the gauging and, hence, the
remaining freedom in the LEEA which then indeed has all the properties
required to model a flop. Only the transition states acquire
a mass away from the transition locus, and 
the potential has a family of degenerate
supersymmetric Minkowski ground states, which is parametrized
by the moduli of $X$.

In order to cope with the technical problems arising in the hypermultiplet sector it is extremely
useful that  {\em every} quaternion-K\"ahler manifold
can be obtained from a so-called hyper-K\"ahler cone by a
superconformal quotient \cite{swann}. As indicated by the name, 
this construction
is intimately related to the construction of hypermultiplet actions
using the superconformal tensor calculus \cite{SCQ,SCgauging,SCQ2}. 
 This treatment does not utilize the fact that the spaces $X(1+N)$ happen to be K\"ahler, but 
 only relies on techniques which 
apply to any quaternion-K\"ahler manifold. Moreover, working on the level of the hyper-K\"ahler cone has the advantage that the product of two hyper-K\"ahler cones is again
a hyper-K\"ahler cone. Thus one can put all the neutral fields
in a separate factor. 
The
isometries of $X(1+N)$ are also obtained from the isometries
of the corresponding hyper-K\"ahler cone. We find the resulting
parametrization very useful for discussing the gauging of the LEEA, as
it is straightforward to see which Killing vector 
corresponds to the gauging describing a flop transition. 
The standard parametrization of $X(1+N)$ \cite{FS}, which relies on its K\"ahler structure, is much less useful for this.

The remaining sections of this paper are organized as follows. 
In section 2 we review the general $\cN = 2, D = 5$ 
gauged supergravity action 
and its relation to CY compactifications of 
M-theory. We explain how the vector multiplet sector can be
determined exactly and introduce an explicit model 
for a flop transition. In section 3 we use the superconformal quotient 
construction to derive the metric and all isometries 
of the unitary Wolf spaces $X(1+N)$. In section 4 we construct an LEEA for 
the specific flop model introduced in subsection 
2.3, which explicitly includes the transition states. 
In section 5 we generalize this setup to a generic flop transition and 
show that our input uniquely fixes the hypermultiplet sector of the 
In-picture Lagrangian. In section 6 we discuss our results and give an outlook on future 
research.

\end{section}

\begin{section}{Five-dimensional Supergravity and Calabi-Yau compactifications}
First we will review the relevant properties of $\cN = 2$, $D = 5$ 
gauged supergravity \cite{VSG,GZ,gs,EGZ}
and its relation to M-theory compactified 
on CY threefolds. For smooth CY compactifications this relation was worked out in  \cite{CY_1}. 
Our conventions for the five-dimensional gauged 
supergravity action follow \cite{UHM}.
We refer to these papers for further details.

\begin{subsection}{Five-dimensional gauged supergravity}
The LEEA of eleven-dimensional supergravity compactified on a smooth 
CY threefold $X$ with Hodge numbers $h^{p,q}$ is given by five-dimensional supergravity coupled to 
$n_V = h^{1,1}-1$ abelian vector and $n_N = h^{2,1}+1 $ neutral 
hypermultiplets. By explicitly including the 
transition states arising in a flop transition we additionally obtain 
$\delta n_H$ charged hypermultiplets. 

The natural starting point for the construction of a LEEA which includes these states is given by the general $\cN = 2, D = 5$ gauged supergravity action
with $n_V$ vector, $n_H = n_N + \delta n_H$ hyper and no tensor 
multiplets. Anticipating the results of sections 4 and 5, we limit ourselves 
to the case of abelian gaugings.
%
The bosonic matter content of this theory consists of the graviton $e_\mu^{~a}$, \mbox{$n_V+1$} vector fields $A^I_\mu$ with field 
strength $F_{\mu \nu}^I = \partial_\mu A^I_\nu - \partial_\nu A^I_\mu$, $n_V$ real vector multiplet scalars $\phi^x$, and $4 
n_H$ real hypermultiplet scalars $q^X$. The bosonic part of the Lagrangian reads: 
\bea\label{2.1}
\nonumber \sqrt{-g}^{-1} \cL^{\cN = 2}_{\rm bosonic} & = &  - \frac{1}{2} R - \frac{1}{4} a_{IJ} F^I_{\mu\nu} F^{J~\mu\nu} 
\\ 
&& - \frac{1}{2} g_{XY} \cD_{\mu} q^X \cD^{\mu} q^Y - \frac{1}{2} g_{xy} \cD_{\mu} \phi^x \cD^{\mu} \phi^y \\ 
\nonumber && + \frac{1}{6 \sqrt{6}} C_{IJK} \sqrt{-g}^{-1} \epsilon^{\mu\nu\rho\sigma\tau} F^I_{\mu \nu} F^J_{\rho \sigma} 
A^K_\tau - {\rm g}^2 \pV(\phi, q) \, .  \eea
The scalars $\phi^x$ and $q^X$ parametrize a 
very special real manifold $\cM_{\mscr{VM}}$ \cite{VSG} and 
a quaternion-K\"ahler\footnote{In parts of the physical literature,
including \cite{BW},  these manifolds are called `quaternionic'. However,
in the mathematical literature quaternionic is a weaker condition
than `quaternion(ic)-K\"ahler'. 
Definitions for both kinds of manifolds are
given later in the main text.}
manifold $\cM_{\mscr{HM}}$ with Ricci scalar
${\cal R}= -8 n_H (n_H +2)$ 
\cite{BW}, respectively.

The vector multiplet sector is determined by the completely symmetric tensor $C_{IJK}$, appearing 
in the Chern-Simons term. This tensor is used to define a real homogeneous
cubic polynomial
\be\label{2.3}
{\cal V}(h) = C_{IJK} \, h^I \, h^J \, h^K \, 
\ee
in $n_V + 1$ real variables $h^I$. The $n_V$-dimensional 
manifold ${\cal M}_{\mscr{VM}}$ 
is obtained by restricting this polynomial to the hypersurface 
\be\label{2.4}
{\cal V}(\phi) = C_{IJK} \, h^I(\phi) \, h^J(\phi) \, h^K(\phi) = 1 \, .
\ee
The coefficients $a_{IJ}$ appearing in the kinetic term of the vector field strength are given by
\be\label{2.5}
\begin{split}
a_{IJ}(h) & := - \, \left. \frac{1}{3} \, \frac{\partial}{\partial h^I} \, \frac{\partial}{\partial h^J} \, \ln {\cal V}(h) 
\right|_{ {\cal V} = 1}  \\
& = -2 \, C_{IJK} \, h^K + 3 \, C_{IKL} \, C_{JMN} \, h^K h^L h^M h^N \, .
\end{split}
\ee
Defining
\be\label{2.6}
h^I_x := - \, \sqrt{\frac{3}{2}} \, \frac{\partial}{\partial \phi^x} \, h^I(\phi) \, , \quad h_I := C_{IJK} h^J h^K \, ,
\ee
the metric on $\cM_{\mscr{VM}}$ is proportional to the pullback\footnote{The $a_{IJ}$
can be interpreted as a metric on the space into which ${\cal M}_{\mscr{VM}}$
is immersed by (\ref{2.4}).}
of $a_{IJ}$,
\be\label{2.7}
g_{xy}(\phi) := h^I_x \, h^J_y \, a_{IJ} \,.
\ee

The hypermultiplet scalars $q^X$ parametrize a quaternion-K\"ahler manifold of dimension $\dim_{\Rom}(\cM_{\mscr{HM}}) = 4 n_H$. 
For $n_H > 1$ such manifolds are characterized by their holonomy group,
\be\label{2.7a}
Hol(\cM_{\mscr{HM}}) = SU(2) \cdot USp(2 n_H) \, ,
\ee
 while in the case $n_H = 1$ they are defined as Einstein spaces
with self-dual Weyl curvature. The restricted holonomy group 
implies that the curvature tensor decomposes into an $SU(2)$ and $USp(2 n_H)$ part
\be\label{2.8}
R_{XYWZ} \, f^W_{iA} \, f^{Z}_{jB} = \epsilon_{ij} \, R_{XY AB}  + C_{AB} \, R_{XYij} \, .
\ee
Here $i=1,2$ is an $SU(2)$ index and 
$A = 1, \ldots, 2 n_H$ is a $USp(2n_H)$ index. These are raised and lowered by the symplectic 
metrics $\epsilon_{ij}$ and $C_{AB}$, respectively. The $4n_{H}$-bein 
$f^{iA}_X$ is related to the metric on $\cM_{\mscr{HM}}$ by
\be\label{2.9}
g_{XY} = f_X^{iA} f^{jB}_Y \epsilon_{ij} C_{AB} = f^{iA}_X \, f_{YiA} \, ,
\ee
and satisfies:
\be\label{2.10}
f^X_{iA} \, f^{iA}_Y = \delta^{~X}_Y \, , \quad f^X_{iA} f^{jB}_X = \delta_i^{~j} \delta_A^{~B} \, .
\ee
Local supersymmetry requires the $SU(2)$ part of the curvature to be non-vanishing \cite{BW}. This feature {\it excludes} hyper-K\"ahler manifolds as 
target manifolds, since these have  trivial $SU(2)$ curvature.

The superconformal quotient construction \cite{SCQ,SCgauging} employed in the next section provides a method to obtain all 
the quantities of interest in the hypermultiplet sector. In this approach the metric $g_{XY}$ and all its isometries are computed from 
the corresponding quantities of the associated hyper-K\"ahler cone, without the need to introduce the vielbein $f^{iA}_X$. 
However, to be able to relate our results to the mayor part of the
literature on hypermultiplets, we review the 
properties of quaternion-K\"ahler  manifolds using the vielbein $f^{iA}_X$. 

We first introduce the Levi-Civita connection $\Gamma_{YZ}^{~~~X}$, a $USp(2n_H)$ connection $\omega_{XB}^{~~~A}$, and an 
$SU(2)$ connection $\omega_{Xk}^{~~~i}$. The vielbein is covariantly constant with respect to these connections,
\be\label{2.11}
\partial_X f^{iA}_Y - \Gamma_{XY}^{~~~Z} f^{iA}_Z + f^{iB}_Y  \omega_{XB}^{~~~A} + \omega_{Xk}^{~~~i} f^{kA}_Y = 0 \, .
\ee
The $SU(2)$ curvature can be expressed in terms of the vielbein as
\be\label{2.12}
\cR_{XYij} = f_{XC(i} \, f^C_{j)Y} \, .
\ee
Raising the index $j$ with $\epsilon^{ij}$, we can expand the $SU(2)$ curvature in terms of the standard Pauli matrices,
\be\label{2.13}
\cR_{XYi}^{~~~~~j} = \imag \, \cR^r_{XY} \, \left( \sigma_r \right)_i^{~j} \, ,
\ee
where $r = 1,2,3$ enumerates the Pauli matrices. The $ \cR^r_{XY} $ defined in this way are real and satisfy
\be\label{2.14}
\cR^r_{XY} \, \cR^{sYZ} = - \, \frac{1}{4} \delta^{rs} \delta_{X}^{~Z} - \frac{1}{2} \, \epsilon^{rst} \cR_{X}^{t~Z} \, .
\ee
It is no accident that the above formula resembles the
quaternionic algebra. A quaternion-K\"ahler manifold is 
in particular quaternionic, i.e., 
there locally exists a triplet of almost complex structures, which satisfy
the quaternionic algebra. The curvatures  $ \cR^r_{XY} $ are
proportional to these almost complex structures. However, since in general none of these almost complex structures is integrable, a quaternion-K\"ahler manifold does not need to be K\"ahler, or, in fact, not even complex.

We now turn to the isometries of ${\cal M}_{\mscr{HM}}$ which are
relevant for the gauging. These must be compatible with 
the three locally defined almost complex structures, i.e., they leave the  
almost complex structures invariant up to an $SU(2)$ rotation. 
Such isometries are called tri-holomorphic.
Given a tri-holomorphic Killing vector $K^X_I(q)$ on the quaternion-K\"ahler manifold, the $ \cR^r_{XY}$ can be used to 
construct an $SU(2)$ triplet of real prepotentials $P^r_I(q)$, the so-called moment maps \cite{mm_1}:\footnote{Up to a 
rescaling, these are identical to the $\hat{\mu}^r$ constructed in the next section.}
\be\label{2.15}
\cR^r_{XY} K^Y_I = D_X \, P^r_I \, ,  \quad D_X P^r_I := \partial_x P^r_I + 2 \epsilon^{rst} \omega^s_X P^t_I \, .
\ee
Here $\omega^s_X$ is defined by $\omega_{Xi}^{~~~j} =: \imag \, \omega^r_X (\sigma_r)^{~j}_i$. Using eq. (\ref{2.14}) this 
relation can be solved for the Killing vector $K^Y_I(q)$:
\be\label{2.16}
K^Z_I = - \frac{4}{3} \cR^{r~ZX} D_X P^r_I \, .
\ee
Hence the moment map $P^r_I$ provides 
a triplet of functions from which the Killing vectors of $\cM_{\mscr{HM}}$ can be obtained. 
%
%
Additionally, one can show that eq. (\ref{2.15}) determines the prepotentials uniquely. In particular, covariantly constant 
shifts $P^{r(0)}_I$ are excluded. This is shown by first contracting eq. (\ref{2.15}) with $D_X$ and then using the harmonicity 
property of the prepotentials \cite{hp}:
\be\label{2.18}
P^r_I = \frac{1}{2 n_H} D_X \left( K_{IY} \cR^{r~XY} \right) \, .
\ee
By virtue of eq. (\ref{2.15}), this relation implies $P^{r(0)}_I = 0$ as $\cR^r_{XY} K^Y_I = 0$ for a covariantly constant 
shift. Hence there is no analog of $D=4$, \mbox{$\cN = 1$} Fayet-Iliopoulos terms in $D=5$, $\cN =2 $ supergravity with a non-trivial 
hypermultiplet sector.

We now discuss the gauging of the Lagrangian (\ref{2.1}).
The scalars take values in a Riemannian manifold, and the gauge
group must operate on them as a subgroup of the isometry group
in order to keep the action invariant.
The procedure which allows one to construct the 
gauge couplings is known as `gauging isometries of the scalar
manifold' \cite{VSG,GZ,gs,EGZ}. This procedure includes the  
covariantization of  the derivatives appearing in the scalar kinetic terms with 
respect to isometries 
of the vector or hypermultiplet target manifolds,
\be\label{2.2}
\cD_{\mu} q^X := \partial_{\mu} q^X + {\rm g} A^I_\mu K^X_{I}(q) \, , \quad 
\cD_{\mu} \phi^x := \partial_{\mu} \phi^x + {\rm g} A^I_\mu K^x_{I}(\phi) \, .
\ee
Here the  $ K^X_{I}(q)$ and $ K^x_{I}(\phi)$ are the Killing vectors of the gauged isometries in the hypermultiplet and 
vector multiplet scalar manifold, respectively. An important consequence of the gauging is that we now have a non-trivial scalar potential $\pV(\phi, q)$. Since we have both vector and hypermultiplets but no tensor multiplets, this potential is determined by the  gauging of hypermultiplet isometries. In order to write down $\pV$ 
explicitly, we define:
\be\label{2.19}
P^r := h^I(\phi) P^r_I \, , \quad P^r_x := h^I_x P^r_I \, , \quad K^X := h^I(\phi) K^X_I \, .
\ee
Here $h^I(\phi)$ are the scalars (\ref{2.3}) associated to the gauge field $A^I_\mu$, $K_I^X(q)$ denotes the Killing 
vector of the hypermultiplet isometry for which $A^I_\mu$ serves as a gauge connection, and $P^r_I$ is its associated $SU(2)$ triplet of moment maps. 
%
%
%
The scalar potential takes the form
\be\label{2.20}
\pV(\phi, q) = -4 P^r P^r + 2 g^{xy} P^r_x P^r_y  + \frac{3}{4} g_{XY} 
K^X K^Y \;.
\ee

Under some conditions \cite{UHM}, this potential 
can be rewritten in terms
of a real function ${\cal W}$. 
This `stability form' is useful, because
it is sufficient to guarantee the gravitational stability of the
theory \cite{stabil}. In five dimensions the relation 
between ${\cal W}$ and $\pV$ is:
\be\label{2.23}
\pV(\phi, q) = -6 \cW^2 + \frac{9}{2} g^{\Lambda \Sigma} \partial_{\Lambda} \cW \partial_{\Sigma} \cW \, .
\ee
Here $\phi^\Lambda$, $\Lambda, \Sigma = 1, \ldots n_V + 4 n_H$ denotes the 
combined set of vector and hypermultiplet 
scalar fields, $g^{\Lambda \Sigma}$ is the direct sum of the  vector and hypermultiplet inverse metrics,
\be\label{4.14}
g^{\Lambda \Sigma}(\phi, q) := g^{XY}(q) \oplus g^{xy}(\phi) \, ,
\ee
and $\cW$ is given by
\be\label{2.21}
\cW := \sqrt{\frac{2}{3} \, P^r \, P^r } \, .
\ee
The equivalence of eqs. (\ref{2.20}) and (\ref{2.23}) requires that when splitting $P^r$ into its norm and phase,
\be\label{2.22}
P^r = \sqrt{\frac{3}{2}} \, \cW \, Q^r \, , \quad Q^r \, Q^r = 1 \, ,
\ee
the phase $Q^r$ is independent of the vector
multiplet scalars, $\partial_x Q^r = 0$. 
As we will show, the scalar potentials of our models indeed satisfy this condition, thereby guaranteeing that the vacua of the theory are stable.

\end{subsection}
\begin{subsection}{Calabi-Yau compactifications}
When compactifying eleven-dimensional supergravity on a smooth CY threefold $X$ \cite{CY_1}, one obtains a five-dimensional ungauged supergravity action, i.e, all 
fields are neutral under the gauge group $U(1)^{n_V + 1}$ and
there is no scalar potential.
In this case the objects introduced above acquire a geometrical interpretation: the vector multiplet scalars encode the deformations of the 
K\"ahler class of $X$ at fixed total volume, while the hypermultiplet
scalars parametrize the volume of $X$, deformations of its 
complex structure, and deformations of the three-form gauge field.
The hypermultiplet containing
the volume is called the universal hypermultiplet, because it is
insensitive to the complex structure of $X$.
Further, the $C_{IJK}$ determining the vector multiplet sector of the LEEA  are given by the triple intersection numbers of $X$,
\be\label{5.4}
C_{IJK} = D_I \cdot D_J \cdot D_K \;,
\ee
where the $D_I$, $I=0,\ldots, n_V$  are a basis of the homological four-cycles
$H_4(X,\mathbbm{Z})$. The  dual basis $C^I$ for the two-cycles is defined by
\be\label{5.3}
C^I \cdot D_J = \delta^I_{~J} \,.
\ee
By integrating the K\"ahler form over the two-cycles $C^I$ we obtain
the quantities
\be\label{5.2}
\hat{h}^I = \int_{C^I} \, J \, ,
\ee
which control the volumes of even-dimensional cycles of $X$.
In particular, the overall volume of $X$ is given by
\be\label{5.6}
vol(X) = \frac{1}{3!} \int_X \, J \wedge J \wedge J = 
\ft16 C_{IJK} \hat{h}^I \hat{h}^J \hat{h}^K \, .
\ee
Since the modulus corresponding to the total volume belongs to the 
universal hypermultiplet, one needs to introduce rescaled 
fields
\be\label{5.7}
 h^I =  (6 \, vol(X))^{-1/3} \, \hat{h}^{I}\, , 
\ee
in order to separate the vector and hypermultiplet moduli \cite{CY_1,flop}. 
These rescaled moduli appear in the cubic  polynomial (\ref{2.4}). 
Moreover, one needs to split the volume into the volume
modulus ${\rm V}$, which is a dynamical field, and a fixed
reference volume ${\rm v}$, which relates the eleven-dimensional 
 and the five-dimensional gravitational couplings,
\be\label{Defv}
vol (X) = {\rm v \cdot V} \;,\;\;\;\mbox{where} \;\;\;
\ft{{\rm v}}{\kappa_{(11)}^2} =  \ft{1}{\kappa_{(5)}^2} \;.
\ee
\end{subsection}
\begin{subsection}{Flop transitions}
We now review the geometry and M-theory
physics of flop transitions. The discussion follows \cite{Witten1}, but
uses the terminology of \cite{Wilson}. Useful 
references for background information are \cite{Huebsch,GreRev,WitD=2}.

The K\"ahler moduli space of a CY threefold $X$ is a cone, called
the K\"ahler cone. The vector multiplet moduli space is 
the projectivization of this cone, or, equivalently, a hypersurface
corresponding to fixed total volume.
At the boundaries of the K\"ahler cone some submanifolds of $X$
contract to zero volume and $X$ becomes a singular CY space 
$\hat{X}$. If there is a second, inequivalent, way to resolve the singularities of $\hat{X}$, leading to a smooth but topologically different CY manifold $\tilde{X}$, then $X$ and $\tilde{X}$ are said to be related by a topological
phase transition. 

We will consider type I
contractions, which lead to flop transitions. In this case the K\"ahler cones 
of $X$ and
$\tilde{X}$ can be glued together along their common boundary
$h^\star = 0$. By joining the K\"ahler cones of all 
CY threefolds related by flops, one obtains the 
extended K\"ahler cone.

In a flop transition
$N$ isolated holomorphic curves ${\cal C}_i$, $i=1,\ldots, N$,
which belong to the same homology class $C^\star=q_I C^I$ and
have volume $h^\star = q_I h^I$, are contracted
to zero volume. By wrapping M2-branes around these holomorphic cycles, one obtains BPS states which carry the charges $\pm (q_I)$, $I=0, \ldots, n_V$ under the
gauge fields $A^I_{\mu}$. This means that these states are charged
under the gauge group $U(1) \subset U(1)^{n_V+1}$, which corresponds to
the gauge field $A_{\mu}^{\star} = q_I A^I_{\mu}$. 
Their masses are
proportional to the volume of the holomorphic curves. By dimensional reduction
of the M2-brane action one computes the mass \cite{flop}
\be
\label{BPS_mass}
M = T_{(2)} ( 6 {\rm v} )^{1/3} q_I h^I \;,
\ee
where $T_{(2)}$ is the tension of the M2-brane.
Since $Z = \pm q_I h^I$ is the central charge of the charged states
with respect to the five-dimensional
supersymmetry algebra, we recognize the five-dimensional BPS mass
formula $M_{(\mscr{BPS})} = \mbox{const} \cdot |Z|$.\footnote{
From the eleven-dimensional point of view the mass of a
wrapped M2-brane is given by $M_{(11)} = T_{(2)} vol ( {\cal C})
= T_{(2)} ( 6 {\rm v V} )^{1/3} q_I h^I$. However, the 
relation between the eleven-dimensional and five-dimensional
metrics involves a conformal rescaling by the volume modulus
${\rm V}$.}

For a flop transition these BPS-states are charged hypermultiplets, i.e., one  obtains $N$
massless charged hypermultiplets at the boundary $h^\star=0$ of the
K\"ahler cone of $X$. As long as the charged hypermultiplets
have finite mass, $h^\star > 0$, physics at scales below this mass can
be described by the effective action $S$ derived from dimensional
reduction on $X$. We call this the Out-picture LEEA, because
it does not contain the transition states. Its vector multiplet
sector is completely determined by the triple intersection
numbers $C_{IJK}$. In the flop transition the curves ${\cal C}_i$
are contracted to zero volume and then re-expanded to
holomorphic curves $\tilde{C}_i$ in the homology class
$\tilde{C}^\star = - C^\star$.  
The triple intersection numbers
$\tilde{C}_{IJK}$ of $\tilde{X}$ are related to those of $X$ 
by
\be
\tilde{C}_{IJK} = C_{IJK} - \frac{N}{6} (D_I \cdot C^{\star})
(D_J \cdot C^{\star})(D_K \cdot C^{\star}) \;.
\label{DeltaC}
\ee
As long as the curves $\tilde{\cal C}_i$ have positive volume,
they support $N$ charged hypermultiplets with finite mass
 proportional to $ \tilde{h}^\star = - h^\star > 0$. For energy scales below this mass
we can use the standard LEEA $\tilde{S}$, obtained by dimensional
reduction on $\tilde{X}$, whose vector multiplet sector is determined
by the $\tilde{C}_{IJK}$. The actions $S$ and $\tilde{S}$ do not
become singular in the limit $h^\star \rightarrow 0$.\footnote{
This is different in four dimensions, where 
the gauge couplings receive
threshold corrections which depend logarithmically on the mass
of the charged states which are integrated out. Therefore 
the actions $S$ and $\tilde{S}$ become singular at the
transition locus and the use of an extended action $\hat{S}$
is indispensable. See \cite{LMZ} for an explicit example.}
Therefore one can in principle avoid to include the extra light modes and
instead consider the coefficients $C_{IJK}$ as piecewise
constant functions, which are discontinuous at the transition locus.

However, a complete low energy description in the vicinity of the boundary requires that we work with an extended LEEA, $\hat{S}$, which contains
the transition states explicitly. The vector multiplet sector
of $\hat{S}$ is completely determined by the coefficients
$\hat{C}_{IJK}$ of its prepotential, which we would like to
determine in terms of the $C_{IJK}$. To find this relation one notes
that there is an intermediate
regime, where the transition states have small, but non-vanishing
masses. Here  both the actions $\hat{S}$ and $S$ (or $\tilde{S}$)  are
valid. Therefore one can relate $\hat{S}$ to $S$ (or  $\tilde{S}$)
by integrating out the transition states. In the vector multiplet sector
this can be done exactly. This result of \cite{Witten1} can be brought to 
a suggestive form by writing the $\hat{C}_{IJK}$ as
`averaged triple intersection numbers' \cite{SU2},
\be
\hat{C}_{IJK} = \ft12 \left( C_{IJK} + \tilde{C}_{IJK} \right) \;,
\label{OSR}
\ee
where $C_{IJK}$ and $\tilde{C}_{IJK}$ are related by 
(\ref{DeltaC}). The change $C_{IJK} \rightarrow \tilde{C}_{IJK}$
 can be viewed as a
threshold effect resulting from integrating in the extra 
hypermultiplets at $h^\star > 0$, continuing to 
$h^\star < 0$ and then integrating them out again. 
We remark that it does not make
sense to use the extended action $\hat{S}$ far away from the flop line, where 
the transition states have a considerable mass, because the full 
M-theory contains many other massive states which are not included in $\hat{S}$. Moreover,
there are additional boundaries of the K\"ahler cones of $X$ and
$\tilde{X}$, where some other states become massless.

So far we have seen that the vector multiplet
sector of $\hat{S}$ can be determined exactly. In the hypermultiplet sector, however,  one has the problem that
it is very hard to compute the quaternion-K\"ahler metric on
${\cal M}_{\mscr{HM}}$ using string or M-theory. The main result
of this paper is that one can find a gauged supergravity action
with $N$ charged hypermultiplets which at least has all the qualitative
properties required of $\hat{S}$. This derivation will be the subject of the 
sections 3 -- 5.

Let us conclude with some remarks. 
For CY threefolds constructed using toric methods, 
the basic two-cycles
$C^I$ can be chosen such that the K\"ahler cone of $X$ is given 
by $h^I > 0$ \cite{Oda}. We will call this the 
`adapted parametrization'.
Then the collapse of a curve at the boundary
$h^J=0$ is described by setting $q_I = \delta_{IJ}$ in the
above formulae. This leads to some 
simplifications, as we have $h^\star = h^J$, $C^{\star} = C^J$, 
$A^{\star}_{\mu} = A^J_{\mu}$, etc. We will not
assume the existence of such a parametrization in our
general discussions, and only use it in the particular
example introduced in the next subsection.

There are also other types of singularities which can
occur at the boundaries of the K\"ahler cone. The only other
cases involving finitely many transition states are the 
type III contractions, which lead to $SU(2)$ 
gauge symmetry enhancement. In this case one also finds
a relation of the form (\ref{OSR}) between the vector multiplet
sectors of the effective actions $S,\hat{S}$ and $\tilde{S}$ \cite{SU2}.
Since for both type I and type III  contractions there
is an underlying $\mathbbm{Z}_2$-action
on the K\"ahler cone, we will refer to 
(\ref{OSR}) as the `orbit sum rule'. Type II contractions
give rise to tensionless strings, which implies that
there are infinitely many additional light states. The so-called cubic cone corresponds
to situations where the volume of $X$ goes to zero. 
In these two cases there are no analogues of $\hat{S}$ in the
framework of five-dimensional gauged supergravity.
Note, however, that when considering type II string theory
on the same CY threefold, these boundaries correspond to
the non-geometric phases.


%
\end{subsection}
\begin{subsection}{The $\mathbbm{F}_1$-model}
We will now consider an explicit example of a CY threefold $X$ with
a flop transition involving a single isolated holomorphic curve.  In this case one charged hypermultiplet becomes 
massless at the transition locus.  
The corresponding CY space is known as the `elliptic fibration over
the first Hirzebruch surface', or $\mathbbm{F}_1$-model for short, and all its relevant properties can
be found in \cite{Jan,phasetransitions}. 
The extended K\"ahler cone of $X$ consists of two K\"ahler cones
which were called regions II and III in \cite{phasetransitions}.
In terms of adapted parametrizations, $h^I > 0$ and $\tilde{h}^I > 0$,
the prepotentials (\ref{2.4}) of these regions are given by \cite{Jan}\footnote{Note that the $t^I$ appearing in 
\cite{Jan,phasetransitions} are related to the $h^I$ by $h^I = 6^{-1/3} t^{I+1}$.}
\be\label{2.24}
\begin{split}
{\cal V}_{II} = & 6 (h^0)^3 + 9 (h^1)^3 + 27 (h^0)^2 h^1 + 27 h^0 (h^1)^2 + 9 (h^0)^2 h^2 + 9 (h^1)^2 h^2  \\
& + 3 h^0 (h^2)^2 + 3 h^1 (h^2)^2 + 18 h^0 h^1 h^2 = 1 \, ,
\end{split}
\ee
and
\be\label{2.25}
{\cal V}_{III} = 8 (\h^0)^3 + 9 (\h^0)^2 \h^1 + 3 \h^0 (\h^1)^2 + 6 (\h^0)^2 \h^2 + 6 \h^0 \h^1 \h^2 = 1 \, ,
\ee
respectively. The transition locus is given by $h^1 \rightarrow 0$ and $\h^2 \rightarrow 0$, respectively.

To analyze the transition it is convenient to introduce variables $T, U, W$, which can be used in both regions. They are given by
\be\label{2.26}
\begin{array}{lll}
6^{1/3} h^0  = W \, , \quad &  6^{1/3} h^1 = U - W \, , \quad &  6^{1/3} h^2 = T - \frac{3}{2} U \, ,  \\
6^{1/3} \h^0  = U \, , \quad & 6^{1/3} \h^1 = T - \frac{1}{2} U - W \, , \quad & 6^{1/3} \h^2 = W - U \, . 
\end{array}
\ee
These formulae also encode the mutual relation between the adapted variables
$h^I$ and $\tilde{h}^I$.
In terms of $T,U,W$, 
the prepotentials (\ref{2.24}) and (\ref{2.25}) become
\be\label{2.27}
\begin{split}
{\cal V}_{II} = & \frac{3}{8} U^3 + \frac{1}{2} U T^2 - \frac{1}{6} W^3 = 1 \, ,\\
{\cal V}_{III} = & \frac{5}{24} U^3 + \frac{1}{2} U^2 W - \frac{1}{2} U W^2 + \frac{1}{2} T^2 U = 1 \, .
\end{split}
\ee
The flop line is located at $U = W$. Comparing these prepotentials, we find that they differ by
\be\label{2.28}
{\cal V}_{II} - {\cal V}_{III} = \frac{1}{6} (U - W)^3 \, .
\ee
This discontinuity in the triple intersection numbers $C_{IJK}$ exactly matches the contribution arising from integrating 
out one charged hypermultiplet \cite{Witten1}.  

We now describe the vector multiplet moduli space corresponding to 
these regions. For this purpose we solve the constraints 
(\ref{2.27}) for $T$, taking $U$ and $W$ as independent scalar fields which parametrize the 
vector multiplet scalar manifolds of the regions II and III. These regions are shown in the first diagram of Fig. \ref{eins}. 
\begin{figure}[t]
\renewcommand{\baselinestretch}{1}
\begin{center}
\leavevmode
\epsfxsize=0.45\textwidth
\epsffile{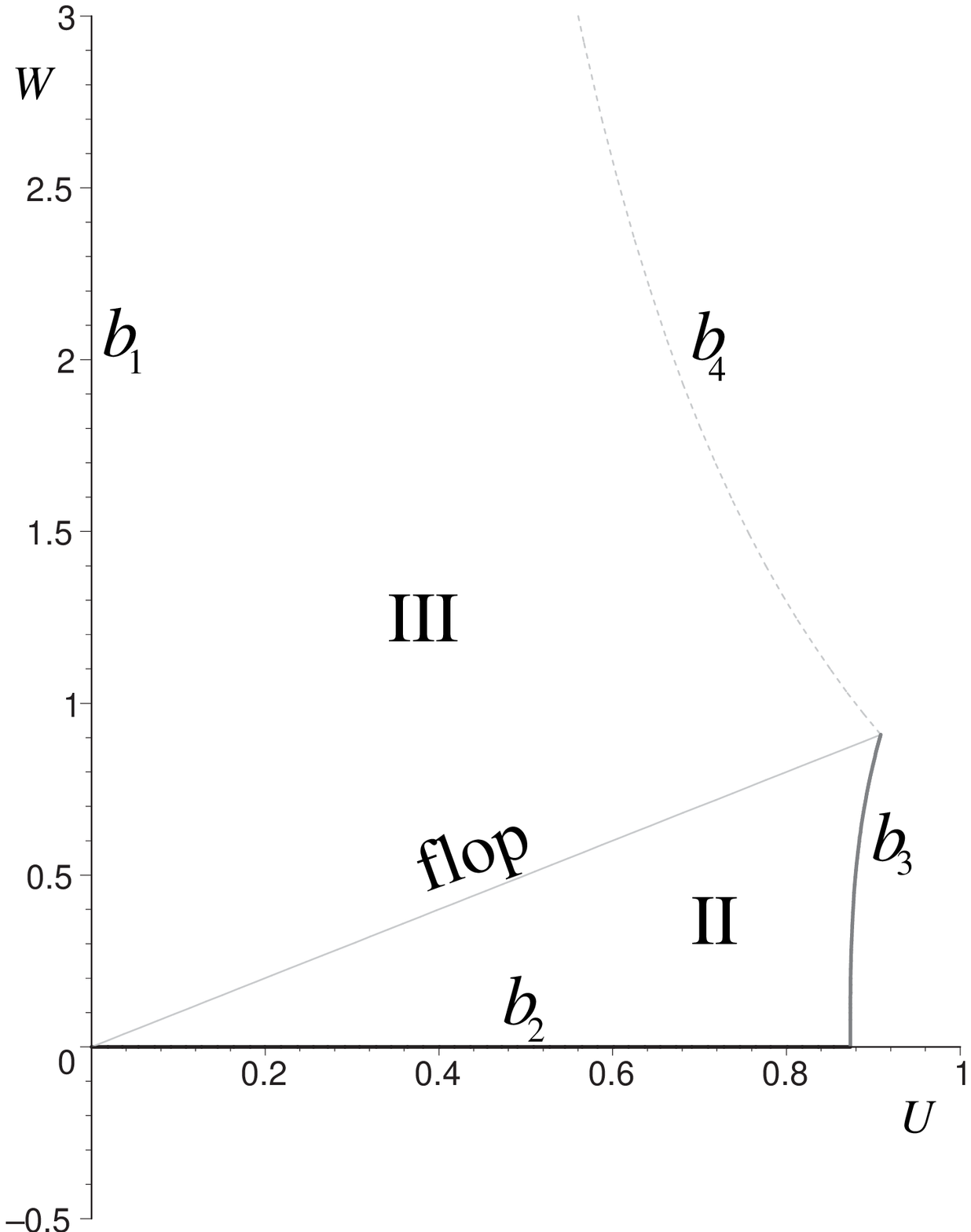} \, \, \, \, \,
\epsfxsize=0.45\textwidth
\epsffile{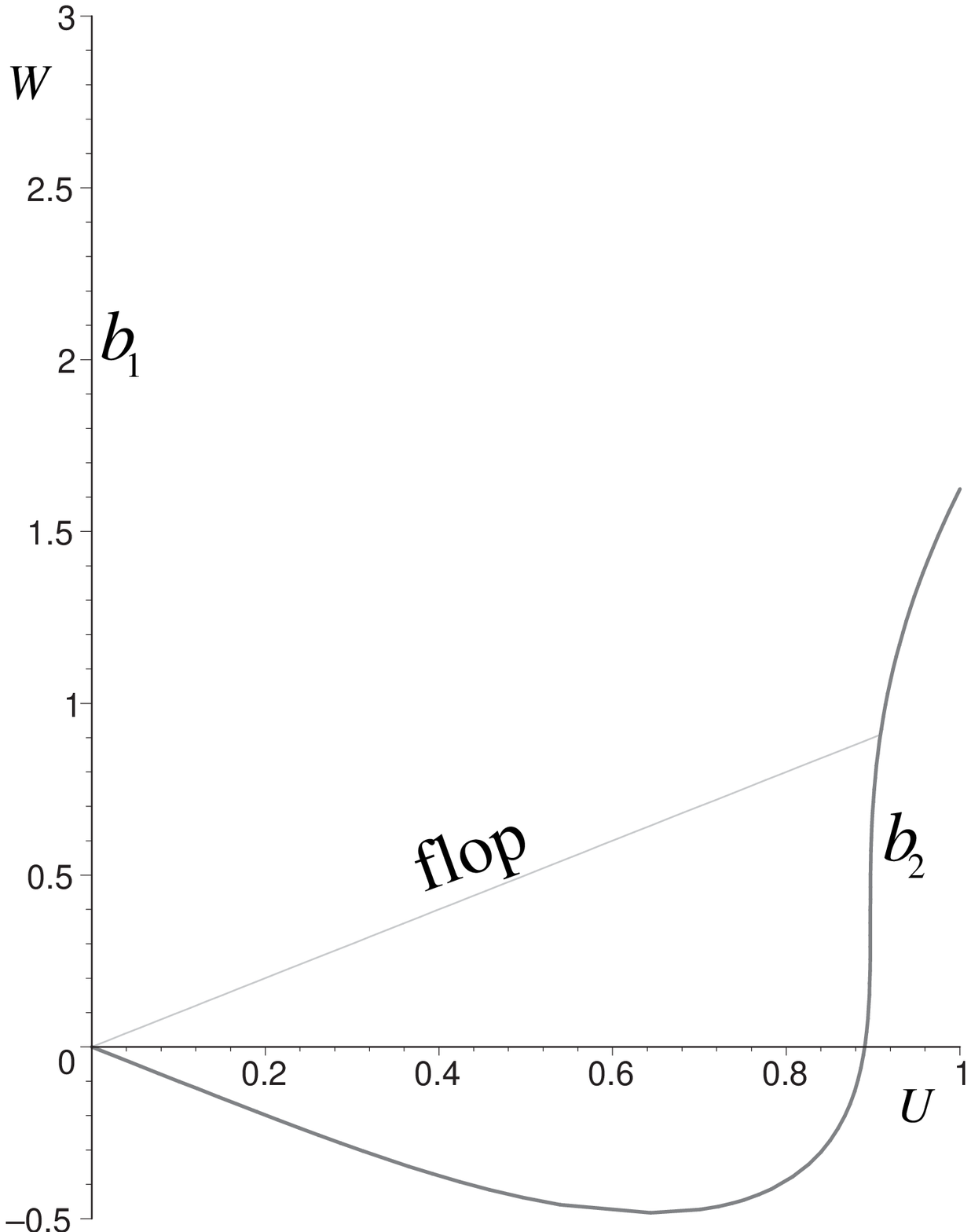}
\end{center}
\parbox[c]{\textwidth}{\caption{\label{eins}{\footnotesize Comparison between the vector multiplet scalar manifolds in the 
Out- (LHS)  and In-picture (RHS). The gray line labeled ``flop'' indicates the locus of the flop transition, $U=W$. 
The ``$b_1$'', ``$b_2$'', ``$b_3$'' and ``$b_4$'' denote the other boundaries of the scalar manifolds. The location of these boundaries is different in the Out- and In-picture.}}}
\end{figure}
Besides the flop line, this diagram displays additional boundaries labeled ``$b_1$'', ``$b_2$'', ``$b_3$'' and 
``$b_4$'', which have the following meaning:
\begin{itemize}
\item The  boundary $b_1$ corresponds to $\h^0 \rightarrow 0$. The metric on the K\"ahler cone has an infinite eigenvalue. In the full K\"ahler cone this limit corresponds to the CY volume becoming zero. However,
the vector multiplet manifold of the five-dimensional supergravity
theory corresponds to a hypersurface of the K\"ahler cone, obtained by
keeping the total volume constant. In this subspace the singularity
takes a different form: while some two-cycles collapse, others diverge,
such that the total volume remains at a fixed finite value \cite{MohProc}.
\item The boundaries $b_2$ and $b_3$ correspond to $h^0 \rightarrow 0$ and $h^2 \rightarrow 0$, respectively. Here the metric on the K\"ahler cone degenerates and has a zero 
eigenvalue. In the microscopic picture a surface is contracted to a 
point and one
obtains tensionless strings. Furthermore, the line $h^0 \rightarrow 0$  is the fixed volume section
of the K\"ahler cone arising from the elliptic fibration over
$\mathbbm{CP}^2$. Since one divisor has been blown down, this space has
one K\"ahler modulus less. This boundary component has been called
region I in \cite{phasetransitions}.
%
%
\item The $b_4$ boundary corresponds to $\h^1 \rightarrow 0$. The metric on the K\"ahler cone is regular. At this boundary 
one obtains $SU(2)$-enhancement.\footnote{This boundary is one of the models where the corresponding In-picture 
Lagrangian has been worked out in \cite{SU2}.}
\end{itemize}

We now construct the vector multiplet scalar manifold for the In-picture Lagrangian. The corresponding prepotential  is 
determined by the orbit sum rule (\ref{OSR}). Taking the average of the prepotentials ${\cal V}_{II}$ and ${\cal V}_{III}$, one finds 
\be\label{2.29}
\begin{split}
\hat{\cal V} = & \frac{1}{2} \left( {\cal V}_{II} + {\cal V}_{III} \right) \\
= & \frac{7}{24} U^3 + \frac{1}{2} U T^2 - \frac{1}{12} W^3 + \frac{1}{4} U^2 W - \frac{1}{4} U W^2 = 1 \, .
\end{split}
\ee
In order to get the metric $g_{xy}$ on the vector multiplet scalar manifold we take $U$ and $W$ as the vector multiplet 
scalar fields: $\phi^x =  U, W $. Solving the constraint for $T$, we obtain
\be\label{2.30}
T(U,W) = \frac{1}{2} \, \left( \frac{24 - 7 U^3 + 2 W^3 - 6 U^2 W + 6 U W^2}{3 U} \right)^{1/2} \, .
\ee
Using the definitions (\ref{2.5}) and (\ref{2.7}) it is straightforward to compute the metric $g_{xy}$ in the 
In-picture,
\be\label{2.32}
 g_{xy}  =  
\left[ \begin{array}{cc} 
 g_{UU} & g_{WU}  \\ g_{WU} & g_{WW}
\end{array} \right] \, .
\ee
Abbreviating
\be\label{2.33a}
K :=  24 - 7 U^3 + 2 W^3 - 6 U^2 W + 6 U W^2 \, ,
\ee
the entries of this matrix are given by
\bea\label{2.33}
\nonumber
g_{UU} & = & \frac{1 }{8 U^2 \, K} \, \bigg( 144 - 6 W^4 U^2 + 48 U W^2 + 4 W^5 U - 72 U^2 W \,  \\
\nonumber && \qquad \qquad  - 14 U^3 W^3 - 17 U^4 W^2 - 168 U^3 + 24 W^3 + W^6 \bigg) \, , \\
 g_{WU} & = &  \frac{1}{8 U \, K} \, \bigg( 24 U W - 36 U^2 - 12 W^2 - W^5 - 4 W^4 U \\
\nonumber && \qquad \qquad  + 6 U^2 W^3 +  14 U^3 W^2 + 17 U^4 W \bigg)  \, ,  \\
\nonumber
g_{WW} & = & \frac{1}{8 \, K} \, \left( 48 (U+W) + 4 U W^3 - 6 U^2 W^2 - 14 U^3 W + W^4 - 17 U^4 \right) \,  .
\eea

The corresponding vector multiplet scalar manifold is shown in the second diagram of Fig. \ref{eins}. Besides the flop line 
at $U = W$ where the metric is regular, this diagram shows two additional boundaries, labeled ``$b_1$'' and ``$b_2$''. 
These have the following meaning:
\begin{itemize}
\item The boundary $b_1$ corresponds to $U \rightarrow 0$. Here the metric (\ref{2.32}) has an infinite eigenvalue.
\item At the boundary $b_2$ the metric $g_{xy}$ degenerates and has a zero eigenvalue.
\end{itemize} 
Comparing the two diagrams shown in Fig. \ref{eins}, we find that the boundaries of the vector multiplet scalar manifolds in 
the Out- and the In-picture are not precisely the same. 
As explained above, the In-picture LEEA can only be expected
to capture the low energy dynamics of M-theory in the vicinity
of the flop line $U=W$. In particular, the dynamics near the other
boundaries of the K\"ahler cone is dominated by other states which 
become light. Therefore it is not clear how to interpret the
behavior of the In-picture LEEA far away  from the flop line and especially at the boundaries in terms of M-theory physics. Nevertheless, the scalar manifold
characterized by 
(\ref{2.32}) and depicted in Fig. \ref{eins} defines a consistent supergravity
action which can be studied in its own right.

\end{subsection}
\end{section}

\begin{section}{The hypermultiplet target manifolds}

Let us now come to our main issue, the construction of a 
family of hypermultiplet target manifolds,
which can be used to describe the transition states 
occurring in a flop transition. In this course we will not attempt to derive these manifolds
directly from M-theory, but use the Wolf spaces (\ref{1.1}).
In order to find the explicit LEEA we need to know
the metrics, the Killing vectors, and the moment maps of these spaces
explicitly. As already mentioned, the Wolf spaces also happen to be K\"ahler, so that one can
derive the metric and the Killing vectors from the
corresponding K\"ahler potential. However, the structure relevant
for the gauging of isometries is the quaternionic structure, as 
the scalar potential depends on the
moment maps of the Killing vectors, which form a 
triplet under the $SU(2)$ related to the 
quaternionic structure.
Therefore
we will construct these objects from the corresponding quantities on  the associated hyper-K\"ahler cone using the superconformal quotient construction \cite{SCQ,SCgauging,SCQ2}. This method
can be applied to any quaternion-K\"ahler 
space. 

The construction of the Wolf spaces (\ref{1.1}) has been
described in \cite{SCQ} but explicit formulae for the metric
have only been given for $X(1)$. The general form of the Killing vectors of $X(N+1)$ has been obtained in \cite{SCgauging}.
In this section we will derive explicit formulae for
the metrics, Killing vectors, and moment maps of all these spaces, while in the next sections
we demonstrate that the resulting parametrization is
extremely useful for including the transition states in the LEEA.

Before considering the particular family (\ref{1.1}) of
quaternion-K\"ahler spaces, let us briefly explain 
the underlying method. 
From the physical point of view the basic idea is to construct
theories with Poincar\'e supersymmetry as gauge-fixed versions
of superconformal theories. In the case at hand one starts with
a theory of $n = N+2$ hypermultiplets\footnote{For notational
convenience we have set $ N+1 = n - 1$.} invariant under rigid 
superconformal transformations. The corresponding
hypermultiplet manifold ${\cal M}_{\mscr{HM}}^{\mscr{SC}}$
is a hyper-K\"ahler cone, i.e., it is hyper-K\"ahler and, in addition, possesses a homothetic
Killing vector $\chi^a$ satisfying $D_a \chi^b = \delta_a^{~b}$.
This implies 
that the hyper-K\"ahler metric $g_{ab}$ of 
${\cal M}_{\mscr{HM}}^{\mscr{SC}}$ has a hyper-K\"ahler potential
$\chi$, with $\chi_a=D_a \chi$ and $g_{ab} = D_a D_b \chi$.
Moreover, ${\cal M}_{\mscr{HM}}^{\mscr{SC}}$ is a cone
over a so-called tri-Sasakian manifold with radial coordinate
$r= \sqrt{2 \chi}$. 
Superconformal invariance also implies that by multiplying the homothety 
$\chi^a$ with the $SU(2)$
triplet of complex structures $\vec{J}= \left[ J^+,J^-,J^3 \right]$ of 
${\cal M}_{\mscr{HM}}^{\mscr{SC}}$ 
one obtains an $SU(2)$ triplet of Killing vectors,
\be\label{3.1a}
\vec{k}^a = \vec{J}^a_{~b} \chi^b \, .
\ee
Using the superconformal calculus, the rigid superconformal theory
can be coupled to conformal supergravity and thus be promoted
to a locally superconformal theory. This theory is
gauge-equivalent to a theory of $n-1$ hypermultiplets
coupled to Poincar\'e supergravity. In this reinterpretation
one of the hypermultiplets becomes dependent on the other
fields and acts as a compensator. Geometrically this
gauging corresponds to performing a superconformal quotient
of ${\cal M}_{\mscr{HM}}^{\mscr{SC}}$ with respect to the
four conformal Killing vector fields  $\chi^a, \vec{k}^a$. 
The resulting hypermultiplet manifold
${\cal M}_{\mscr{HM}}$ of the Poincar\'e supergravity theory
is quaternion-K\"ahler. In fact every quaternion-K\"ahler
manifold can be obtained by this construction from its
associated hyper-K\"ahler cone \cite{swann}.

The construction of the Wolf spaces $X(n-1)$ which have
dimension $\dim_\Rom(X(n-1)) = 4 (n-1)$ proceeds in several steps. First one needs to obtain the 
hyper-K\"ahler cone $\cH^{(2n)}$ associated with 
the space $X(n-1)$. In \cite{SCQ} this cone has been constructed as the 
hyper-K\"ahler quotient \cite{HC} of flat 
$\C^{(2n+2)}$ with respect to a particular tri-holomorphic $U(1)$ isometry.
Then the superconformal quotient  is taken in two steps. First one quotients  
$\cH^{(2n)}$ by the homothetic
Killing vector $\chi^a$ and the Killing vector $k^3=[k^{3a}]$ corresponding
to the Cartan direction of the $SU(2)$ isometry group.
This quotient is a standard K\"ahler quotient \cite{KQ}. 
The resulting space is 
the twistor space $\cZ^{(2n-2)}$ over $X(n-1)$. In the second step one 
quotients $\cZ^{(2n-2)}$ by the remaining Killing 
vectors $k^+$ and $k^-$. The isometry (\ref{3.1a}), 
however, is only holomorphic and not tri-holomorphic. This implies that 
at the level of the twistor space, $k^+$ and $k^-$ are isometries up to 
$SU(2)$ rotations. In order to obtain well 
defined quantities on the quaternion-K\"ahler manifold 
one has to include a compensating $SU(2)$ transformation.
For convenience we have summarized all spaces appearing in this construction in Table \ref{t.1}.
\begin{table}[t]
\begin{tabular*}{\textwidth}{@{\extracolsep{\fill}} cccc} \hline \hline
symbol & space &  $\dim_\Rom$ & coordinates \\ \hline
$\C^{(2n+2)}$ & flat complex space & $4 n + 4$ & $ \left\{z_+^I, \zb_+^I, z_{-I}, \zb_{-I} \right\}$  \\ 
$\cH^{(2n)}$ & HKC over $X(n-1)$ & $ 4n$ & $\left\{z_+^{\prime~ a}, \zb_+^{\prime~a}, z_{-a}^{\prime}, \zb_{-a}^{\prime} 
\right\}$  \\
$\cZ^{(2n-1)}$ & twistor space & $ 4 n-2 $ & 
$\left\{v^i , \vb^i , u_i, \ub_i , \zeta, \bar{\zeta}  \right\}$  \\ 
$X(n-1)$ &  Wolf space & $4n-4$ & $\left\{v^i , \vb^i , u_i, \ub_i  \right\}$  \\
\hline \hline
\end{tabular*}
\renewcommand{\baselinestretch}{1}
\parbox[c]{\textwidth}{\caption{\label{t.1}{\footnotesize
Summary of the spaces appearing in the quotient construction of the unitary Wolf spaces $X(n-1)$. The index ranges are given by 
$I = 1 , \ldots , n+1$ , $a = 1 , \ldots , n$ and $i = 1 , \ldots , n-1$.  }}}
\end{table}  

The rest of this section is organized as follows. In subsection 3.1 we start with flat $\C^{(2n+2)}$ and construct the 
metric on $X(n-1)$. The result is given in eqs. (\ref{3.30}) and (\ref{3.31}). In subsection 3.2 we construct the Killing vectors of this 
metric and their tri-holomorphic moment maps. These are given in eq. (\ref{3.42}) and (\ref{3.45}), respectively. In 
subsection 3.3 we use these general results to explicitly construct the Cartan
subgroup of the isometry group of $X(2)$. This is needed to construct the gauged  
LEEA for the model introduced in subsection 2.3. In subsection 3.4 we establish 
the relation between the quantities constructed in this section and the conventions of the gauged supergravity Lagrangian 
(\ref{2.1}). A reader not interested in the technical details of this section may adopt the main results and directly 
proceed to section 4.

\begin{subsection}{The metrics of the Wolf spaces $X(n-1)$}
\subsubsection*{The starting point: flat complex space $\C^{(2n+2)}$}
We start our construction by considering $\C^{(2n+2)}$ with complex coordinates $z^I_+, z_{-I}$, where $I = 1, \ldots, n+1$. 
The metric is taken to be K\"ahler with the K\"ahler potential
\be\label{3.2}
\chi_{(2n+2)} := \eta_{I\Jb} \, \zp^{I} \, \zbp^{\Jb} + \eta^{I\Jb} \, z_{-I} \, \bar{z}_{-\Jb} \, .
\ee
Here $\eta_{I \Jb} = \mbox{diag}\left[ - , \ldots , - , + , +\right]$ has indefinite signature $(p,q)$ with $p = n-1$ negative 
and $q = 2$ positive eigenvalues, and $\eta^{I \Jb}$ is its inverse. This signature of $\eta$ ensures that the space 
obtained from the quotient construction is of non-compact type, as required by supergravity. Later on the coordinates associated 
with the negative eigenvalues of $\eta$ will play the role of the hypermultiplet scalars, while the coordinates with 
the positive eigenvalues act as gauge compensators.

We now promote $\C^{(2n+2)}$ to a hyper-K\"ahler manifold. For this purpose we introduce the coordinates $x^{\tI} := 
\left\{ z_+^I, \bar{z}_+^{\bar{I}}, z_{-I}, \bar{z}_{-\bar{I}}\right\}$, with $\tI = 1, \ldots, 4n+4$. With 
respect to these coordinates, the $SU(2)$ triplet of complex structures is taken to be:
\bea\label{3.3}
\nonumber J^{1 \tI}_{~~ \tJ} & = & \left[ \begin{array}{cccc} 0 & 0 & 0 & \unit \\ 0 & 0 & \unit & 0 \\ 0 & - \unit & 0 & 0 
\\ - \unit & 0 &0 & 0   \end{array} 
\right] \, , \quad
J^{2 \tI}_{~~ \tJ} = \left[ \begin{array}{cccc} 0 & 0 & 0 & \imag \unit \\ 0 & 0 & - \imag \unit & 0 \\ 0 & - \imag \unit & 
0 & 0 \\ \imag \unit & 0 &0 & 0   \end{array} 
\right], \\
J^{3 \tI}_{~~ \tJ} & = & \left[ \begin{array}{cccc} \imag \unit & 0 & 0 & 0 \\ 0 & - \imag \unit & 0 & 0 \\ 0 & 0 & \imag 
\unit & 0 \\ 0 & 0 &0 & - \imag \unit  \end{array} 
\right]. 
\eea
Here the entries are $(n+1) \times (n+1)$ dimensional block matrices and $\unit$ denotes the corresponding unit matrix. 
These complex structures satisfy the quaternionic algebra $J^r J^s = - \unit \delta^{rs} + \epsilon^{rst} J^t$, with $r,s,t 
= 1,2,3$ being the $SU(2)$ index. The complex coordinates $ z_+^I, \bar{z}_+^{\bar{I}}, z_{-I}$ and $\bar{z}_{-\bar{I}}$ are 
defined with respect to the canonical complex structure $J^3$. The K\"ahler metric $g_{\tI \tJ}$ derived from (\ref{3.2}) is 
hermitian with respect to all three complex structures, $g(J^r \, \cdot, J^r \, \cdot) = g( \cdot \, , \, \cdot)$. 

Instead of working with the basis (\ref{3.3}), it is more convenient to use $J^{\pm} := \frac{1}{2} 
\left( J^1 \pm \imag J^2 \right)$, since quantities defined with respect to this basis will turn out to be (anti-)holomorphic 
with respect to $J^3$. 
From these complex structures we obtain the following $SU(2)$ triplet of K\"ahler forms:
\bea\label{3.4}
\nonumber \omega^3 & = & - \imag \, \eta_{I\Jb} \, \rmd \zp^{I} \, \wedge \, \rmd \zbp^{\Jb} - \imag \, \eta^{I\Jb} \, \rmd 
z_{-I} \, \wedge \, \rmd \bar{z}_{-\Jb} \, , \\
\omega^+ & = & \rmd \zp^I \, \wedge \, \rmd z_{-I} \, , \qquad \omega^- = \bar{\omega}^+ \, . 
\eea
Their components are given by   
\be\label{3.5}
\Omega^3_{\tI \tJ} = g_{ \tI \tK} \, J^{3 \tK}_{~~~\tJ} \, , \quad \Omega^+_{\tI \tJ} = g_{\tI \tK} \, J^{+ \tK}_{~~~\tJ} 
\quad \mbox{and} \quad \Omega^-_{\tI \tK} = \bar{\Omega}^+_{\tI \tK}  \, ,
\ee
respectively. Here the ``bar'' denotes complex conjugation with respect to $J^3$.

Let us now consider the {\it linear} action of the $U(n-1, 2)$ isometry subgroup:
\be\label{3.6}
 z^I_+  \rightarrow  U^I_{~J} \, z^J_+ \, , \qquad 
z_{-I}  \rightarrow  \left( U^{-1} \right)^J_{~I} \, z_{-J} \,. 
\ee
With respect to this isometry, the $z_+^I$ coordinates transform in the fundamental representation of $U(n-1,2)$, while the 
$z_{-I}$ transform in the complex conjugate representation. Using $\bar{U}^{\bar{I}}_{~\bar{J}} = \eta^{\bar{I}K} 
\left( U^{-1} \right)^{N}_{~K} \, \eta_{N \bar{J}}$ one finds that the K\"ahler potential (\ref{3.2}) is invariant under 
this transformation. In principle the isometry group of (\ref{3.2}) contains additional generators. But since these do not descend to tri-holomorphic isometries of the hyper-K\"ahler cone $\cH^{(2n)}$ they do not give rise to isometries of $X(n-1)$,\footnote{The coset formulation of $X(n-1)$ indicates that the full 
isometry group of $X(n-1)$ is given by $SU(n-1,2)$. In our approach this $SU(n-1,2)$ arises from the $U(n-1,2)$ above modulo 
the $U(1)$ gauged in the hyper-K\"ahler quotient.} and will not be considered here.\footnote{The fact that only tri-holomorphic isometries give rise to isometries of the quaternion-K\"ahler space has been observed in \cite{SCQ}.} 

The Killing vectors of the linearized isometries are given by
\be\label{3.7}
k_{+ \alpha}^I = \imag \, t^{~I}_{\alpha~J} \, z_{+}^J \, , \qquad 
k_{- \alpha I} = - \imag \, t^{~J}_{\alpha~I} \, z_{-J} \, .
\ee
Here $\alpha$ numerates the  $ n(n+2)+1 $ generators of $U(n-1,2)$, $t^{~I}_{\alpha~J} $.  To simplify our 
notation we will drop the index $\alpha$ in the following. The action of these Killing vectors is tri-holomorphic, i.e., the 
Lie derivative with respect to $k$ satisfies $\cL_{k} J^r = 0$ for all three complex structures (\ref{3.3}). This  implies in 
particular that the Killing vectors are holomorphic with respect to $J^3$. Hence we can obtain their components with 
respect to $\bar{z}_+^{\bar{I}}$ and $\bar{z}_{- \bar{I}}$ by complex conjugation of $k^I_+$ and $k_{-I}$, respectively.

The condition that the vectors $k$ are Killing, $\cL_k g = 0$, as well as tri-holomorphic implies that they are  
Hamiltonian, $\cL_k \omega^r = 0$. The last statement provides the integrability condition for the moment maps associated 
with these isometries. They are obtained as the solution of the equation
\be\label{3.7a}
 \frac{\partial \mu^r}{\partial x^{\tI}}  := \Omega^{r}_{\tI \tJ} \, k^{\tJ} \, ,
\ee
 where $r = +,-,3$. Substituting the Killing 
vectors (\ref{3.7}) and the K\"ahler forms (\ref{3.5}), these equations are easily integrated and yield:
\bea\label{3.8}
\nonumber \mu^3 &=& - \bar{z}^{\Jb}_+ \, \eta_{\Jb I} \, t^{I}_{~K} \, z^K_+ \, + \, z_{-I} \, t^{I}_{~J} \, \eta^{J 
\bar{K}} \, \bar{z}_{- \bar{K}} \, , \\
\mu^+ &=& - \imag \, z_{-I} \, t^{I}_{~J} \, z^J_+ \, ,  \qquad \, \mu^- = \bar{\mu}^+ \, .
\eea
Here we omitted the constants of integration, which, in principle, could give rise to Fayet-Iliopoulous terms. Since these 
terms appear in neither the superconformal theory defined on the level of the hyper-K\"ahler cone nor in the $D=5$ 
supergravity action reviewed in subsection 2.1, the moment maps (\ref{3.8}) will give rise to the most general moment maps 
compatible with the action (\ref{2.1}).
\subsubsection*{The hyper-K\"ahler quotient construction of $\cH^{(2n)}$}
We now perform the hyper-K\"ahler quotient construction of $\cH^{(2n)}$ by taking the quotient of $\C^{(2n+2)}$ with respect 
to the $U(1)$ isometry which acts on $z_+^I$ and $z_{-I}$ by opposite phase transformations. The  infinitesimal generator 
of this isometry is given by $t^{I}_{~J} = \delta^{I}_{~J}$. Substituting this generator into (\ref{3.8}) we find:
\be\label{3.9}
 \mu^3 = - \eta_{I \Jb} \, z^I_+ \, \bar{z}^{\Jb}_+  \, + \, \eta^{I \bar{J}} \, z_{-I} \,  \bar{z}_{- \bar{J}}  \, , \quad
\mu^{+} = - \imag \, z^{I}_{+} \, z_{- I} \, , \quad  \mu^- = \bar{\mu}^+ \, .
\ee
The quotient is performed by first introducing $U(1)$ invariant coordinates $z^{\prime~I}_+$, $z^{\prime}_{-~I}$ on 
$\C^{(2n+2)}$ and substituting these into the moment maps (\ref{3.9}). We then set the resulting moment maps to zero and solve 
these constraints in terms of $z^{\prime~n+1}_{+}$, $z_{-n+1}^{\prime}$ and their complex conjugates. The remaining 
unconstrained coordinates $z^{\prime~a}_{+}$, $z_{-a}^{\prime}$  with $a = 1, \ldots ,n$ provide coordinates on 
$\cH^{(2n)}$. 
In practice, we choose the primed coordinates as
\be\label{3.10}
z^{\prime I}_+ := \frac{z^{I}_{+}}{z^{n+1}_+}, \qquad \qquad z^{\prime}_{-I} := z_{-I} \, z_{+}^{n+1} \, .
\ee
In terms of these coordinates the moment maps (\ref{3.9}) become
\bea\label{3.11}
\nonumber \mu^3 &=& - \eta_{I \Jb} \, z^{\prime I}_+ \, \bar{z}^{\prime \Jb}_+  \, \left( \bar{z}^{n+1}_+ \, z^{n+1}_+ 
\right) + \, \eta^{I \bar{J}} \, z_{-I}^{\prime} \,  \bar{z}_{- \bar{J}}^{\prime} \, \left( \bar{z}^{n+1}_+ \, z^{n+1}_+ 
\right)^{-1}  \, , \\
\mu^{+} &=& - \imag \, z^{\prime I}_{+} \, z_{- I}^{\prime} \, , \, \qquad \, \mu^- = \bar{\mu}^+ \, .
\eea
Setting the moment maps to zero and solving the resulting constraints in terms of $z^{\prime~n+1}_{+}$, 
$z_{-n+1}^{\prime}$ and their complex conjugates yields
\be\label{3.12}
\left( \bar{z}^{n+1}_+ \, z^{n+1}_+ \right)  =  \left( \frac{ \eta^{I \bar{J}} \, z_{-I}^{\prime} \,  \bar{z}_{- 
\bar{J}}^{\prime} }{ \eta_{I \Jb} \, z^{\prime I}_+ \, \bar{z}^{\prime \Jb}_+  } \right)^{1/2} \, , \quad
z^{\prime~n+1}_+  =  1 \, , \quad z^{\prime}_{- n+1} = - \, z^{\prime~a}_{+} \, z^{\prime}_{- a} \, .
\ee

Substituting the new coordinates (\ref{3.10}) into the K\"ahler potential (\ref{3.2}) and performing the gauge fixing 
gives the K\"ahler potential for the metric on $\cH^{(2n)}$:
\be\label{3.15}
\chi_{\cH}(z^a_+, \zb_{+}^{a}, z_{-a}, z_{- a}) = 2 \, \chi_+ \, \chi_- \, .
\ee
Here we introduced
\be\label{3.14}
\chi_+ :=  \left( \eta_{I \Jb} \, z^{\prime~I}_+ \, \bar{z}^{\prime~\Jb}_+ \right)^{1/2}  
\, , \qquad \, \chi_- :=  \left( \eta^{I \bar{J}} \, z_{-I}^{\prime} \,  \bar{z}_{- \bar{J}}^{\prime} \right)^{1/2} \, ,
\ee
where it is understood that we have performed the gauge fixing (\ref{3.12}).

In view of the later steps in the construction we also calculate $\omega^+_{\cH}$. Substituting the primed coordinates  into (\ref{3.4}) and performing the gauge fixing gives 
\be\label{3.15a}
\omega^+_{\cH} = \rmd z^{\prime~a}_+ \, \wedge \, \rmd z^{\prime}_{-a} \, .
\ee
\subsubsection*{The superconformal quotient: Going to twistor space $\cZ^{(2n-1)}$}
We now descend to the twistor space $\cZ^{(2n-1)}$. Here we follow \cite{SCQ} and introduce the coordinates
\be\label{3.16}
z^{\prime}_{-i} = \e^{2 z} \,  u_i \, , \qquad z_{-n}^{\prime} = \e^{2 z} \, , \qquad i = 1, \ldots , n-1 \, . 
\ee
We next single out another coordinate, $\zeta$, which will be gauged when going to $X(n-1)$. To this end, we 
substitute the coordinates (\ref{3.16}) into $\omega^+_{\cH}$ given in eq. (\ref{3.15a}):
\be\label{3.17}
\omega^+_\cZ = \e^{2z} \left( 2 \, \rmd z^{\prime~n}_+ + 2 \, u_i \, \rmd z^{\prime~i}_+ \right) \wedge \, \rmd z + 
\e^{2z} \, \rmd z^{\prime~i}_+ \, \wedge \, \rmd u_i \, . 
\ee
Following the general construction of the superconformal quotient, the components of this 2-form should be compared to
\be\label{3.18}
\Omega_{\tilde{a} \tilde{b}}^+\left(u,z,z^{\prime}_+ \right)  = \e^{2z} \, \left[ \begin{array}{cc} \omega_{\tilde{\imath} 
\tilde{\jmath}}(u, z^{\prime}_+ ) & X_{\tilde{\imath}}(u, z^{\prime}_+ ) \\ - X_{\tilde{\jmath}}(u, z^{\prime}_+ ) & 0 
\end{array} \right] \, . 
\ee
From this comparison, we obtain the explicit form of $X_{\tilde{\imath}} $:
\be\label{3.19}
X_{\tilde{\imath}} = \left[ \underbrace{0 \, , \, \ldots \, , \, 0}_{(n-1) \mbox{times}}, \, 2 \, u_1 \, , \,  \ldots \, , 
\, 2 \, u_{n-1} \, , \, 2 \,  \right]^{\rm T} \, .
\ee
We then determine $\zeta$ by first finding a $Y^{\tilde{\imath}}$, subject to $X_{{\tilde{\imath}}} 
Y^{{\tilde{\imath}}} = 1$ and independent of the coordinates $ z^{\prime~a}_+ , u_i, z $. The coordinate $\zeta$ is  
obtained as the solution of the differential equation
\be\label{3.20}
Y^{\tilde{\imath}} \, \frac{\partial}{\partial z^{\tilde{\imath}}} = \frac{\partial}{\partial \zeta} \, .
\ee
Choosing $Y^{\tilde{\imath}} = \left[ 0, \ldots , 0, 1/2 \right]$, which is natural but not unique, we find $\zeta = 2 
z^{\prime~n}_+$. This motivates to introduce
\be\label{3.21}
z^{\prime~i}_+ = v^i \, , \qquad z^{\prime~n}_{+} = \frac{1}{2} \, \zeta \, .
\ee
The $v^i, u_i, \zeta$ and their complex conjugates then provide coordinates on the twistor space $\cZ^{(2n-1)}$. In order to 
obtain the K\"ahler potential of $\cZ^{(2n-1)}$ we first substitute these new coordinates into $\chi_+$ and $\chi_-$:
\bea\label{3.22}
\nonumber \chi_+ & = & \left( 1 + \eta_{i \bar{\jmath}} \, v^i \bar{v}^{\bar{\jmath}} + \frac{1}{4} \, \zeta \, \bar{\zeta} 
\right)^{1/2} \, , \\
\chi_- & = & \e^{z + \bar{z}} \left( 1 + \eta^{i \bar{\jmath}} \, u_i \bar{u}_{\bar{\jmath}} + \left( v^i u_i + \frac{1}{2} 
\zeta \right) \left( \bar{v}^{\bar{\imath}} \, \bar{u}_{\bar{\imath}} + \frac{1}{2} \bar{\zeta}\right)  \right)^{1/2} \, .
\eea

The K\"ahler potential of $\cZ^{(2n-1)}$, $K(v,u,\zeta,\bar{v},\bar{u},\bar{\zeta})$, can be deduced by comparing 
$\chi_{\cH}$ given in (\ref{3.15}) to the following expression:
\be\label{3.23}
\chi_{\cH} = \e^{z + \bar{z} + K(v,u,\zeta,\bar{v},\bar{u},\bar{\zeta})} \, .
\ee
From this we read off
\be\label{3.23a}
K(v,u,\zeta,\bar{v},\bar{u},\bar{\zeta}) = \ln \left( \chi_+ \right) + \ln \left( \chi_- \right) + \ln \left( 2 \right) \, ,
\ee
where $\chi_+$ and $\chi_-$ are taken at $z = 0$.

In order to calculate the $SU(2)$ compensators appearing in the construction of the metric of $X(n-1)$, we also need 
$\omega_{\cZ}^+$ in terms of the coordinates $ v^i, u_i, z , \zeta $. By substituting these coordinates into (\ref{3.17}) we 
obtain
\be\label{3.27}
\omega^+_\cZ = \e^{2z} \left( \rmd \zeta + 2 \, u_i \, \rmd v^i \right) \wedge \rmd z + \e^{2z} \, \rmd v^i \, \wedge \, 
\rmd u_i \, .
\ee
\subsubsection*{The superconformal quotient: The metric on $X(n-1)$}
We now descend to the quaternion-K\"ahler space $X(n-1)$ by setting $\zeta =  0$. The K\"ahler potential $K$ 
becomes
\be\label{3.24}
K(u,v, 0,\bar{u},\bar{v}, 0) = \frac{1}{2} \, \ln\left( \phi_+ \right) + \frac{1}{2} \, \ln\left( \phi_- \right) +  
\ln\left( 2 \right) \, ,
\ee
where we introduced
\be\label{3.25}
 \phi_+ := 1 + \eta_{i \bar{\jmath}} \, v^i \bar{v}^{\bar{\jmath}}  \, , \quad 
\phi_- :=  1 + \eta^{i \bar{\jmath}} \, u_i \bar{u}_{\bar{\jmath}} + \left( v^i u_i \right) \left( \bar{v}^{\bar{\imath}} \, 
\bar{u}_{\bar{\imath}} \right)  \, .
\ee
However, since $\zeta$ is not parallel to the Killing vector $k^+$ given in (\ref{3.1a}), the condition  $\zeta = 0$ is 
not  preserved. In order to obtain the metric $G_{\alpha \bar{\beta}}$ on $X(n-1)$ we need to include an additional 
compensating transformation. Explicitly we have
\be\label{3.26}
G_{\alpha \bar{\beta}} = K_{\alpha \bar{\beta}} - \e^{-2 K} \, X_{\alpha} \, \bar{X}_{\bar{\beta}} \, ,
\ee
where $K_{\alpha \bar{\beta}}$ is the K\"ahler metric obtained from (\ref{3.24}) and $\alpha = 1, \ldots , 2n-2$ enumerates 
the coordinates $\{v^1,\ldots,v^{n-1}, u_1, \ldots, u_{n-1} \}$. In order to determine the explicit from of the $X_{\alpha}$ 
appearing in the compensating transformation, we compare the components of $\omega^+_\cZ$ given in (\ref{3.27}) to the 
general form of $\omega^+_{\cZ}$  given in \cite{SCQ}:
\be\label{3.28}
 \Omega^+_{ab} = \e^{2z} \left[ \begin{array}{ccc} \omega_{\alpha \beta}(v,u) & 0 & X_{\alpha}(v,u) \\ 0 & 0 & 1 \\ - 
X_{\beta}(v,u) & -1 & 0 \end{array} \right] \, .
\ee
From this we read off
\be\label{3.29}
X_{\alpha} = \left[ \, 2 \, u_1 \, , \, \ldots \, , \, 2 \, u_{n-1} \, , \underbrace{ \, 0 \, , \, \ldots \, , 0 \, }_{(n-1) 
\mbox{ times} } \right]^{\rm T} \, .
\ee

Having all these ingredients at hand, we can now write down the metric 
(\ref{3.26}) explicitly. Arranging our indices as $X,Y = \left\{ v^i, \vb^{\ib}, u_i, \ub_{\ib} \right\}$ the components of $G_{\alpha \bar{\beta}}$ can be read off from the following matrix:
\be\label{3.30}
 G_{XY}  = \left[ \begin{array}{cccc}
 0 & G_{v \vb} & 0 & G_{v \ub} \\ 
 G_{\vb v} & 0 & G_{\vb u} & 0 \\
 0 & G_{u \vb} & 0 & G_{u \ub} \\
 G_{\ub v} & 0 & G_{\ub u} & 0 \\
\end{array} \right] \, .
\ee
The entries of this matrix are given by
\bea\label{3.31}
\nonumber G_{u_i \ub_{ \jb }}  & = & \frac{1}{2 \phi_-}  \left( \eta^{i \jb} + \vb^{\jb} \, v^i \right) 
- \frac{1}{2 \phi_-^2}
\left( \eta^{\jb l} u_l + \vb^{\jb} \left( v^l u_l \right) \right)  
\big( \eta^{i \lb} \ub_{\lb} + v^i \big( \vb^{\lb} \ub_{\lb} \big) \big) \, , \\
G_{\vb^{\ib} u_{j}} & = &  \frac{1}{2 \phi_-^2} \, \left( \ub_{\ib} v^j \, \left( 1 + \eta^{k \lb} u_k \ub_{\lb} \right) - 
\ub_{\ib} \eta^{j \lb} \ub_{\lb} \left( v^l u_l \right) \right)  \, , \\
\nonumber G_{v^i \vb^{\jb}} & = & 
\frac{1}{2 \phi_+} \, \eta_{i \jb} 
- \frac{1}{2 \phi_+^2} \, \big( \eta_{i \lb} \, \vb^{\lb} \big) \, \big( \eta_{\jb l} \, v^{l} \big) \, - \, \frac{1}{\phi_+ 
\phi_-} \, u_i \ub_{\jb} \\
\nonumber && + \, \frac{1}{2 \phi_-} \, u_i \, \ub_{\jb} \, - \, \frac{1}{2 \phi_-^2} \, u_{i} \, \ub_{\jb} \, \big( v^l \, 
u_l \big) \big( \vb^{\lb} \ub_{\lb} \big) \, .
\eea
The other non-vanishing entries of the matrix can be obtained from the relations $G_{v^i \vb^{\jb}} = G_{\vb^{\jb} v^i}$, $G_{u_i \ub_{\jb}} = G_{\ub_{\jb} u_i}$, $G_{v^i \ub_{\jb}} = G_{\ub_{\jb} v^i}$ and $G_{v^i \ub_{\jb}} = \left( G_{\vb^{\ib} u_j} \right)^*$, where ``${}^*$'' denotes complex conjugation.

These results provide metrics of $X(n-1)$, which obviously are hermitian but not K\"ahler with respect to $J^3$. In fact the 
holomorphic assignments in (\ref{3.31}) are adapted to the quaternionic structure, which cannot be used to define a K\"ahler 
potential. However, there must be a non-holomorphic coordinate transformation which brings the metric given above into its 
standard K\"ahler form \cite{FS}.

To conclude this subsection, let us comment on the special case $n=2$, which corresponds to the universal hypermultiplet. In 
this case the index $i$ has only a single value and may be omitted. Setting $\eta_{i \jb} = \eta_{1 \bar{1}}= -1$ the 
general metric (\ref{3.31}) simplifies to
\bea\label{3.32}
\nonumber G_{u \ub} & = & - \, \frac{1}{2 \phi_-^2} \, \left( 1 - v \vb \right) \, ,  \\
G_{u \vb} & = & \frac{1}{2 \phi_-^2} \, \ub v \, , \\
\nonumber G_{v \vb} & = & - \, \frac{1}{2 \, \phi_{+}^{2} \, \phi_-^2} \, \left(1 - u \ub \left( 1 - v \vb \right)^2 \right) 
 \, .
\eea
This is exactly the metric for the universal hypermultiplet derived in \cite{SCQ}.
\end{subsection}
\begin{subsection}{The isometries of the Wolf spaces $X(n-1)$}
After obtaining the metric on $X(n-1)$, we will now derive the second ingredient needed in the construction of the LEEA and 
derive the Killing vectors and moment maps of the unitary Wolf spaces. We follow the calculation of \cite{SCgauging} and extend 
these results.

The Killing vectors of flat $\C^{(2n+2)}$ are given in (\ref{3.7}). In order to find the Killing vectors on 
the hyper-K\"ahler cone $\cH^{(2n)}$ we perform a coordinate transformation to the primed coordinates (\ref{3.10}). The 
resulting Killing vectors read:
\be\label{3.33}
k^{\prime a}_{+}  =  \imag \, t^a_{~I} \, z^{\prime I}_+ - \,  \imag  \, z^{\prime a}_{+} \, t^{n+1}_{~~I} \,  z^{\prime 
I}_{+} \, , \quad 
k^{\prime}_{- a}  =  \, \imag \, z_{-a}^{\prime} \, t^{n+1}_{~~~I} \, z^{\prime~I}_+ - \, \imag \, t^{I}_{~a} \, 
z^{\prime}_{-I} \, .
\ee
Here we have implicitly performed the gauge fixing (\ref{3.12}).

To obtain the Killing vectors on $X(n-1)$ we first transform (\ref{3.33}) into the coordinates $ v^i, u_i, z, \zeta$ given by
\be\label{3.34}
z^{\prime~i}_+ = v^i \, , \qquad z^{\prime~n}_+ = \frac{1}{2} \, \zeta \, , \qquad z_{-i}^\prime = \e^{2 z} \, u_i \, , 
\qquad z_{-n}^{\prime} = \e^{2 z} \, .
\ee
Fixing $\zeta = 0$, the resulting vectors $k^\alpha$ read:
\be\label{3.35}
\begin{split}
k^{v^i} & = \, \imag \, t^i_{~j} \, v^j + \imag \, t^i_{~n+1} - \imag \, v^i \, t^{n+1}_{~~~j} \, v^j - \imag \, v^i \, 
t^{n+1}_{~~~n+1} \, , \\
k^{\zeta} & = 2 \, \imag \, \left( t^n_{~i} \, v^i + t^n_{~n+1} \right) \, , \\
k_z & = \frac{\imag}{2} \left( t^{n+1}_{~~~i} v^i + t^{n+1}_{~~~n+1} - t^{i}_{~n} \, u_i - t^n_{~n} + t^{n+1}_{~~~n} \, v^i 
\, u_i \right) \, , \\
k_{u_i} & = \imag \, u_i \, \left( t^{n+1}_{~~~j} \, v^j + t^{n+1}_{~~~n+1} \right) - \imag \, t^j_{~i} u_j - \imag \, 
t^{n}_{~i} + \imag \, t^{n+1}_{~~~i} \,\left( v^j \, u_j \right) - 2 \, u_i \, k_z \, .
\end{split}
\ee
However, these vectors do not preserve the gauge $\zeta = 0$. In order to get the Killing vectors $\hat{k}^\alpha$ on 
$X(n-1)$ we have to implement an additional compensating transformation, which is given by \cite{SCgauging}:
\be\label{3.36}
\kh^\alpha = k^\alpha - \, \frac{X^\alpha}{X^\zeta} \, k^\zeta \, .
\ee
According to \cite{SCQ}, $X^\alpha, X^\zeta$ can be determined from the equations
\be\label{3.37}
X^\alpha = \left( \hat{\omega}^{\alpha \beta} K_\beta + Z^\alpha \, K_\zeta \right) \, \e^{2 K} 
\, , \quad 
X^{\zeta} = \left( 1 - Z^{\alpha} K_\alpha \right) \e^{2 K} \, ,
\ee
with $\hat{\omega}^{\alpha \beta}$ and $Z^\alpha$ given by
\be\label{3.38}
\hat{\omega}^{\alpha \gamma} \omega_{\gamma \beta} = - \delta^\alpha_{~\beta} 
\, , 
\qquad Z^{\alpha} = - \, \hat{\omega}^{\alpha \beta} \, X_{\beta} \, .
\ee
Here $K$ is the K\"ahler potential (\ref{3.23a}), $K_\alpha$ and $K_{\zeta}$ denote its derivative with respect to $ v^i, u_i 
$ and $\zeta$, respectively, and $X_\alpha$ is given in (\ref{3.29}). The $\omega_{\alpha \beta}$ is determined by 
comparing $\omega^+_\cZ$ given in (\ref{3.27}) with the general expression (\ref{3.28}) and $\hat{\omega}^{\alpha \beta}$ is obtained from eq. (\ref{3.38}). Explicitly, we find
\be\label{3.39}
\omega_{\alpha \beta} = \left[ \begin{array}{cc} 0 & \unit \\ - \unit & 0 \end{array} \right] 
\, , \qquad 
\hat{\omega}^{\alpha \beta} = \left[ \begin{array}{cc} 0 & \unit \\ - \unit & 0 \end{array} \right] \, ,
\ee
where $\unit$ denotes the $n-1$-dimensional unit matrix. Substituting $X_\alpha$ into (\ref{3.38}) gives
\be\label{3.40}
Z^{\alpha} = \left[ \underbrace{ \, 0 \, , \ldots , \, 0}_{n-1 \, \mbox{times}} \,  , \,  2 \, u_1 \, , \,  \ldots \, , \, 2 
\, u_{n-1} \right]^{\rm T} \, .
\ee
With these results at hand, it is now straightforward to write down the explicit form of the compensating transformation 
appearing in (\ref{3.36}):
\be\label{3.41}
\frac{X^{v^i}}{X^{\zeta}} = \frac{1}{2} \left( \eta^{i \jb} \, \ub_{\jb} + v^i \left( \vb^{\jb} \ub_{\jb} \right) \right) \, 
, \qquad 
\frac{X^{u_i}}{X^{\zeta}} = \, - \, \frac{ \phi_{-}}{2 \phi_+} \, \eta_{i \jb} \, \vb^{\jb} \, .
\ee
The Killing vectors of $X(n-1)$ then read:
\bea\label{3.42}
\nonumber \kh^{v^i} & = & \imag \, t^{i}_{~j} \, v^j + \imag \, t^{i}_{~n+1} - \imag \, v^{i} \, t^{n+1}_{~~~j} \, v^{j} - 
\imag \, v^{i} \, t^{n+1}_{~~~n+1} - \, \frac{ k^{\zeta} }{2} \left( \eta^{i \jb} \ub_{\jb} + v^{i} \, \vb^{\kb} \ub_{\kb} 
\right) \, ,\\
\kh^{u_i} & = & \imag \, u_i \, \left( t^{n+1}_{~~~j} v^j + t^{n+1}_{~~~n+1} \right) - \imag \, t^{j}_{~i} \, u_{j} - \imag 
\, t^{n}_{~i} + \imag \, t^{n+1}_{~~~i} \, \left( v^j \, u_j \right) \, \\
\nonumber && \qquad \qquad  - 2 u_i \, k_z + \frac{\phi_-}{2 \phi_+} \, k^{\zeta} \, \eta_{i \jb} \, \vb^{\jb} \, .
\eea
Here $k^{\zeta}$ and $k_z$ are given in (\ref{3.35}).

We will now derive the moment maps associated with these Killing vectors, starting from the moment maps on flat 
$\C^{(2n+2)}$ given in (\ref{3.8}). Rewriting them in terms of the primed coordinates, the corresponding moment 
maps on $\cH^{(2n)}$ are:
\be\label{3.43}
\begin{split}
\mu^3 & = - \, \frac{2}{\chi_{(2n)}} \left( \chi_-^2 \, \zb_+^{\prime \Ib} \, \eta_{\Ib J} t^J_{~K} \, z^{\prime K}_+ - 
\chi_+^2 \, z^{\prime}_{-K} \, t^K_{~J} \, \eta^{J \Ib} \zb^{\prime}_{- \Ib} \right) \, , \\
\mu^+ & = - \, \imag \, z^{\prime}_{- I} \, t^{I}_{~J} \, z^{\prime J}_+ \, , \qquad \mu^- = \bar{\mu}^+ \, .
\end{split}
\ee
Again, it is understood that these expressions implicitly contain the gauge fixing (\ref{3.12}).

In \cite{SCgauging} it was found that the moment maps on the hyper-K\"ahler cone, $\mu^r$, and the moment maps on the underlying 
quaternion-K\"ahler manifold, $\hat{\mu}^r$, are related by
\be\label{3.44}
\mu^3 = \chi_{(2n)} \, \hat{\mu}^3 \, , \qquad \mu^+ = \e^{z - \zb} \, \chi_{(2n)} \, \hat{\mu}^+ \, , \qquad \hat{\mu}^- = 
\bar{\hat{\mu}}^+ \, .
\ee
Substituting in the coordinate transformation (\ref{3.34}) and gauging $\zeta = 0$, we obtain the following expression for the 
moment maps on $X(n-1)$:
\bea\label{3.45}
\hat{\mu}^3 & = & - \, \frac{1}{2 \phi_+} \left\{ \vb^{\ib} \, \eta_{\ib j} \, t^j_{~k} \, v^k + t^{n+1}_{~~~i} \, v^i + 
\vb^{\jb} \, \eta_{\jb i} \, t^i_{~n+1} + t^{n+1}_{~~~n+1} \right\} \\ \nonumber
&&  + \frac{1}{2 \phi_-} 
\big\{  
u_i \, t^i_{~j} \, \eta^{j \kb} \, \ub_{\kb} + u_i \, t^{i}_{~n} + t^n_{~i} \, \eta^{i \jb} \, \ub_{\jb} + t^n_{~n} 
- \left( u_i \, t^i_{~n+1} + t^n_{~n+1}\right) \left( \ub_{\ib} \vb^{\ib} \right) 
\\ \nonumber
&& \qquad \qquad 
- \left( t^{n+1}_{~~~i} \, \eta^{i \jb} \, \ub_{\jb} + t^{n+1}_{~~~n} \right)
 \, \left( u_k v^k \right) 
+ t^{n+1}_{~~~n+1} \, \left( u_i \, v^i \right) 
\, \left( \ub_{\jb} \vb^{\jb} \right)
\big\}  \, , 
\\ \nonumber
\hat{\mu}^+ & = & - \, \frac{\imag}{2 \phi_+^{1/2} \phi_-^{1/2}} \, \big\{ u_i \, t^i_{~j} \, v^j + t^n_{~j} \, v^j - 
t^{n+1}_{~~~i} \, v^i \left( u_j \, v^j \right) + u_i \, t^i_{~n+1} \\ \nonumber
&& \qquad \qquad \qquad + t^n_{~n+1} - t^{n+1}_{~~~n+1} \left(u_i \, v^i \right) \big\} \, .
\eea
This result completes the derivation of the Killing vectors and moment maps of $X(n-1)$. Together with the metric 
(\ref{3.30}) we now have all the ingredients for modeling the hypermultiplet sector of our In-picture LEEA. 
\end{subsection}
\begin{subsection}{Examples of Isometries on $X(2)$}
Before we embark upon this construction, we will use our general results (\ref{3.42}) and (\ref{3.45}) to explicitly 
calculate the Killing vectors and moment maps of the Cartan subgroup of the isometry group on $X(2)$, $SU(2,2)$.  As it will 
turn out in the next section, this information already suffices to construct the In-picture Lagrangian for the model 
introduced in subsection 2.3.

We choose the three Cartan generators of $SU(2,2)$ as
\bea\label{3.46}
\nonumber
C_{1}  & = & \frac{1}{2} \, {\rm diag} \left[ \, 1 \, , \, -1 \, , \, 0 \, , 0 \, \right] \, , \\ 
C_{2} & = & \frac{1}{2} \, {\rm diag}
\left[ 
 \, 0 \, , \, 0 \, , \, 1 \, , -1 \, 
\right] \, , \\
\nonumber
C_{3} & = & \frac{1}{2 \sqrt{2}} \, {\rm diag} 
 \left[ 
 \, 1 \, , \, 1 \, , \, -1 \, , -1 \,
\right]  \, .
\eea
Substituting these matrices into the expression for a generic Killing vector on $X(2)$ (\ref{3.42}), we find:
\bea\label{3.47}
\nonumber
k_1 & = & \frac{\imag}{2} \, \left[ v^1 , - v^2 , - \vb^1 , \vb^2, - u_1, u_2, \ub_1, - \ub_2 \right]^{\rm T} \, , \\
k_2 & = & \frac{\imag}{2} \, \left[ v^1 ,  v^2 , - \vb^1 , - \vb^2,  u_1, u_2, - \ub_1, - \ub_2 \right]^{\rm T} \, , \\
\nonumber 
k_3 & = & \frac{\imag}{\sqrt{2}} \, \left[ v^1 , v^2 , - \vb^1 , -\vb^2, - u_1, - u_2, \ub_1, \ub_2 \right]^{\rm T} \, .
\eea
Here the index $\alpha$ in $k_\alpha$ enumerates the Cartan generators. The components of the Killing vectors are given with 
respect to the basis
\be\label{3.47a}
 \left\{ \partial_{v^1}, 
\partial_{v^2},\partial_{\vb^1},\partial_{\vb^2},\partial_{u_1},\partial_{u_2},\partial_{\ub_1},\partial_{\ub_2} \right\} \, 
.
\ee

When gauging these isometries, we also need the triplet of moment maps corresponding to the Killing vectors. These are 
obtained by evaluating eq. (\ref{3.45}) for the generators (\ref{3.46}). The resulting moment maps are:
\bea\label{3.48}
\nonumber
\hat{\mu}_1 & = & 
\left[ 
\begin{array}{c}
-  \frac{\imag}{4 \phi_+^{1/2} \phi_-^{1/2}}  \, \left( v^1 u_1 - v^2 u_2 - \vb^1 \ub_1 + \vb^2 \ub_2 \right) \\
- \, \frac{1}{4 \phi_+^{1/2} \phi_-^{1/2}}  \, \left( v^1 u_1 - v^2 u_2 + \vb^1 \ub_1 - \vb^2 \ub_2 \right) \\
 \frac{1}{4 \, \phi_+} \left(  \vb^1 v^1 - \vb^2 v^2 \right) - \frac{1}{4 \, \phi_-} \left( \ub_1 u_1 - \ub_2 u_2 \right) 
\end{array}
\right] \, , \\ \nonumber
\hat{\mu}_2 & = & 
\left[ 
\begin{array}{c}
-  \,  \frac{\imag}{4 \phi_+^{1/2} \phi_-^{1/2}} \,  \left( v^1 u_1 + v^2 u_2 - \vb^1 \ub_1 - \vb^2 \ub_2 \right) \\
- \,  \frac{1}{4 \phi_+^{1/2} \phi_-^{1/2}} \, \left( v^1 u_1 + v^2 u_2 + \vb^1 \ub_1 + \vb^2 \ub_2 \right) \\
 \frac{1}{4 \, \phi_+}  + \frac{1}{4 \, \phi_-} \left( 1 - \left( v^1 u_1 + v^2 u_2 \right) \left( \vb^1 \ub_1 + \vb^2 \ub_2 
\right)  \right) 
\end{array}
\right] \, , \\
\hat{\mu}_3 & = &   
\left[ 
\begin{array}{c}
- \,  \frac{ \imag \sqrt{2} }{4  \phi_+^{1/2} \phi_-^{1/2}}  \, \left( v^1 u_1 + v^2 u_2 - \vb^1 \ub_1 - \vb^2 \ub_2 \right) 
\\
- \,  \frac{ \sqrt{2} }{4  \phi_+^{1/2} \phi_-^{1/2}} \left( v^1 u_1 + v^2 u_2 + \vb^1 \ub_1 + \vb^2 \ub_2 \right) \\
 \frac{\sqrt{2}}{4 \phi_+} \left( \vb^1 v^1 + \vb^2 v^2 \right) 
 - \frac{\sqrt{2}}{4 \phi_-} \left( \ub_1 u_1 + \ub_2 u_2  \right) 
\end{array}
\right] \, .
\eea
Here the index $\alpha$ in $\hat{\mu}_\alpha$ again enumerates the Cartan generators. The components of the moment maps are 
given with respect to the basis $\left\{ \hat{\mu}^1 , \hat{\mu}^2, \hat{\mu}^3 \right\}$ associated with the complex 
structures given in (\ref{3.3}). Their relation to $\hat{\mu}^+$ and $\hat{\mu}^-$ is given by
\be\label{3.49}
\hat{\mu}^1 = \hat{\mu}^+ + \hat{\mu}^- \, , \quad \hat{\mu}^2 = - \imag \left(  \hat{\mu}^+ - \hat{\mu}^-\right) \, , \quad 
 \hat{\mu}^3 = \hat{\mu}^3 \, .
\ee
The results (\ref{3.47}) and (\ref{3.48}) complete this section on isometries in the two hypermultiplet case. 
\end{subsection}
\begin{subsection}{The relation to the supergravity conventions}
Matching the conventions given in \cite{UHM} and \cite{SCQ} for the universal hypermultiplet, we find 
 that the metric $G_{XY}$ given in (\ref{3.30}) and the metric $g_{XY}$ in 
the Lagrangian (\ref{2.1})  are related by
\be\label{3.50}
g_{XY}(q) = - G_{XY}(q) \, .
\ee
Looking at the definitions of the moment map (\ref{2.15}) and the one given in \cite{SCgauging}, we further find that 
these differ by a factor of one half,
\be\label{3.51}
P^r_I(q) = \frac{1}{2} \hat{\mu}^r(q) \, .
\ee
\end{subsection}
\end{section}
\begin{section}{The Lagrangian of the $\mathbbm{F}_1$-model}
Now we have all the ingredients to construct the In-picture LEEA for the explicit model introduced in subsection 2.3. We proceed by first deriving the Lagrangian and then showing that the scalar masses obey the conditions arising from the 
microscopic picture. This example already illustrates all the key features that appear in the analysis of a generic 
flop transition in section 5.
\begin{subsection}{Gauging the general supergravity action}
According to the microscopic description of the flop transition reviewed in subsection 2.3 our 
In-picture Lagrangian should contain one neutral and one charged hypermultiplet. These play the roles of the universal 
hypermultiplet and of the transition states, respectively.\footnote{As explained in the introduction, our model only includes the 
universal hypermultiplet and the transition states. The additional neutral hypermultiplets arising in the CY 
compactification are frozen and will not be included in the following analysis.} The latter are charged with respect to the 
vector field $A^\star_\mu$ whose associated cycle collapses at the flop. In our particular model this implies that the 
transition states are charged with respect to the vector field associated with the scalar field combination $\left( U - W 
\right)$, as this is the modulus that vanishes at the transition locus. Taking the hypermultiplet scalar manifold $\cM_{\mscr{HM}}$ 
to be $X(2)$ with complex coordinates  $v^1, v^2, u_1, u_2$, we choose the universal hypermultiplet as being represented by 
$v^1, u_1$, while the transition states are given by $v^2, u_2$. 
 
Our first task is to identify the proper gauging in the hypermultiplet sector. Here we need a $U(1)$ Killing vector which 
acts on the second hypermultiplet only.  To identify this Killing vector we use the results of section 3.3, where the 
$U(1)$ Killing vectors on $X(2)$ have been worked out. Looking at eq. (\ref{3.47}) it turns out that there is (up to 
rescaling\footnote{Any rescaling of the Killing vector can
be absorbed by a rescaling of the gauge coupling ${\rm g}$. We will fix
the normalization of the Killing vector and show later that ${\rm g}$
is uniquely determined by microscopic M-theory physics.}) 
a  {\it unique} linear combination of Killing vectors which is independent of $v^1, u_1$ and does not act on the 
universal hypermultiplet: 
\be\label{4.1}
\begin{split}
k_{\rm gauge} := & k_1 - \, \frac{1}{\sqrt{2}} \, k_3 \\
= & - \, \imag \left[ \, 0 \, , \, v^2 \, , \, 0 \, , \,  - \vb^2 \, , \,  0 \, , \,  - u_2 \,  , \, 0, \,  \ub_2 \, 
\right]^{\rm T} \, . 
\end{split}
\ee
 Taking proper linear combinations of 
the Cartan generators (\ref{3.46}) we find the generator of this isometry is
\be\label{4.2}
C_{\rm gauge} := C_1 - \, \frac{1}{\sqrt{2}} \, C_3 = \frac{1}{4} \, {\rm diag}
\left[ 
\, 1 \, , \, -3 \, , \, 1 \, , \, 1 \, 
\right] \, .
\ee

In order to construct the scalar potential (\ref{2.20}) we also need the triplet of moment maps associated with this 
isometry. These can be derived by either substituting the generator (\ref{4.2}) into the general formula for the  moment map 
(\ref{3.45}) or by taking appropriate linear combinations of the moment maps given in (\ref{3.48}). Using the definition of 
$\phi_+$ and $\phi_-$ (\ref{3.25}) to simplify $\hat{\mu}^3_{\rm gauge}$ we obtain 
\be\label{4.3}
\begin{split}
 \hat{\mu}^r_{\rm gauge} :=  \hat{\mu}_1^r - \, \frac{1}{\sqrt{2}} \, \hat{\mu}^r_3 
 =    \left[ 
\begin{array}{c}
 \frac{\imag}{2 \phi_+^{1/2} \phi_-^{1/2}}  \left(  v^2 u_2 -  \vb^2 \ub_2 \right) \\
 \frac{1}{2 \phi_+^{1/2} \phi_-^{1/2}} \, \left(  v^2 u_2 +  \vb^2 \ub_2 \right) \\
 \frac{1}{2 \phi_-}
  \left(  \ub_2 u_2  \right)
 - \frac{1}{2 \phi_+ }  \left( \vb^2 v^2 \right)
\end{array}
\right] \, .
\end{split}
\ee

In the next step we perform the gauging of (\ref{2.1}) with respect to this isometry. In order to compare our five-dimensional supergravity action with eleven-dimensional M-theory data, it is natural to use the embedding coordinates $h^I$, (\ref{2.24}), as these are the coordinates which are related to the volumes of the CY cycles. However, for the $\mathbbm{F}_1$-model it is more convenient to work with the variables $T,U,W$ given in (\ref{2.26}). Further, it is useful to label 
the vector fields $A^I_\mu$ by their corresponding scalar field:
\be\label{4.4}
\left\{ A^I_\mu \, , \quad I = 0,1,2 \right\} \quad \longrightarrow \quad \left\{A^T_{\mu} \, , \;  A^U_\mu \, , \; A^W_\mu   
\right\} \, .
\ee

Next we consider the scalar kinetic terms of (\ref{2.1}). Since we do not gauge any isometries of the vector multiplet 
scalar manifold, the corresponding gauge covariant derivative becomes a partial derivative,
\be\label{4.5}
\cD_{\mu} \phi^x = \partial_\mu \phi^x \quad \Leftrightarrow \quad K^x_{I}(\phi) \quad \mbox{\rm 
trivial} \, .
\ee
In the hypermultiplet sector the microscopic picture fixes the $U(1)$  gauge connection of the isometry (\ref{4.1}) to be 
$A^U_\mu - A^W_\mu$. To implement this requirement we set
\be\label{4.6}
 K^X_T(q) = 0  \, , \quad K^X_U(q) = k_{\rm gauge}^X(q) \, , \qquad K^X_W(q) = - k_{\rm gauge}^X(q) \, ,  
\ee
where $k_{\rm gauge}^X(q)$ is given in (\ref{4.1}). The covariant derivative for the hypermultiplet scalars then becomes
\be\label{4.7}
\cD_\mu \, q^X = \partial_\mu \, q^X + {\rm g} \left( A^U_\mu - A^W_\mu \right) k_{\rm gauge}^X(q) \, 
. 
\ee
This expression explicitly shows that the universal hypermultiplet parametrized by $v^1, u_1$ is neutral, while the 
transition states $v^2, u_2$ carry $U(1)$ charges
$q = -1$ and $q = +1$  with respect to the gauge fields, respectively.

Next we turn to the scalar potential (\ref{2.20}) where we take the independent vector multiplet scalar fields as $\phi^x =  U, W $, while $T(U,W)$ is given in 
eq. (\ref{2.30}). Including the rescaling (\ref{3.51}) the $P^r_I$ are given by %
\be\label{4.9}
P^r_T = 0 \, , \quad P^r_U = \frac{1}{2} \, \hat{\mu}^r_{\rm gauge} \, , \quad P^r_W = - \, \frac{1}{2} \hat{\mu}^r_{\rm gauge}   
 \, .
\ee
Correspondingly, $P^r$ is obtained as
\be\label{4.10}
P^r = \frac{1}{2} \, h^1  \, \hat{\mu}^r_{\rm gauge} 
= \frac{1}{2} \, 6^{-1/3} \left( U - W \right) \, \hat{\mu}^r_{\rm gauge} \, ,
\ee
where we used (\ref{2.26}) in the second step.

In order to construct the scalar potential of our theory, we work out the superpotential (\ref{2.21}). For the $P^r$ above this is given by
\be\label{4.11}
\cW = 6^{-5/6} \, \left( \frac{1}{2 \phi_-} \, \left( \ub_2 u_2 \right) + \frac{1}{2 \phi_+} \left( \vb^2 v^2 
\right) \right)  \, \left( U - W \right) \, .
\ee
It is now straightforward to check that the $Q^r$ defined in (\ref{2.22}) is independent of the vector multiplet scalars,
\be\label{4.12}
Q^r = \sqrt{\frac{2}{3}} \, \frac{P^r}{\cW} = \frac{\hat{\mu}^r_{\rm gauge}(q)}{\frac{1}{2 \phi_-} \, \left( \ub_2 u_2 
\right) + \frac{1}{2 \phi_+} \left( \vb^2 v^2 \right)} \, .
\ee
Hence the condition $\partial_x Q^r = 0$ is trivially satisfied. This implies that the scalar potential can be written as
\be\label{4.13}
\pV(\phi, q) = -6 \cW^2 + \frac{9}{2} g^{\Lambda \Sigma} \partial_{\Lambda} \cW \partial_{\Sigma} \cW \, .
\ee
Here $g^{\Lambda \Sigma}$ is defined in (\ref{4.14}) and the coordinates of the scalar manifold $\cM_{\mscr{HM}} \otimes \cM_{\mscr{VM}}$ 
are taken to be
\be\label{4.15}
\phi^\Lambda = \left\{ \, v^1 \, , \, v^2 \, , \, \vb^1 \, , \, \vb^2 \, , \, u_1 \, , \, u_2 \, , \, \ub_1 \, , \, \ub_2 \, 
, \, U \, , \, W \,  \right\} \, .
\ee

Alternatively, we can compute the scalar potential by substituting the quantities $K^X$, $P^r$, $P^r_x$, and the (inverse) 
metrics (\ref{2.32}) and (\ref{3.30}) directly into the scalar potential (\ref{2.20}). By explicit computation one finds 
that the resulting expressions agree. Since the equality of (\ref{2.23}) and (\ref{2.20}) requires some non-trivial 
identities of quaternion-K\"ahler geometry, this result provides a non-trivial check for our derivation.
\end{subsection}
\begin{subsection}{Vacua and mass matrix}
After constructing the In-picture Lagrangian for our flop model, let us investigate its 
vacuum structure and calculate the corresponding mass matrix. From the microscopic analysis we know that the 
masses of the transition states must be proportional to $|U-W|$ while all other fields must be massless. 

We start by investigating the critical points of the potential, which are given by the condition $\partial_\Lambda \pV = 0$. 
The expression for the potential (\ref{4.13}) reveals that  all critical points of the 
superpotential $\cW$ are automatically critical points of $\pV$ while the converse is generally not true.
From the explicit form of $\cW$, (\ref{4.11}), one recognizes that $\cW$ consists of terms proportional to $ |u_2|^2$ and $ |v^2|^2$. Hence taking a 
derivative with respect to any scalar field $\phi^\Lambda$ and afterwards setting $v^2 = u_2 = 0$ satisfies the condition 
for $\cW$ having a critical point:
\be\label{4.16}
\partial_\Lambda \, \cW \left|_{v^2 = u_2  = 0} \right. = 0 \, .
\ee
This implies that we have an entire manifold $\cM_C$ of critical points which is parametrized by the vacuum expectation 
values of the universal hypermultiplet and vector multiplet scalars:
\be\label{4.17}
\partial_\Lambda \pV \left|_{\cM_C} \right. = 0 \, , \quad \cM_C = \left\{ 
\begin{array}{l} 
v^2  = u_2 = 0 \\
v^1 , u_1 , U, W \quad \mbox{not determined by eq. (\ref{4.16})} \, .
\end{array} 
\right.
\ee
Corollary 3 from \cite{Vicente} implies that $\cM_C$ actually contains all supersymmetric critical points of $\pV$. At first sight $\pV$ seems to have also some other critical points, but it turns out that these are all located outside the scalar manifolds.

To determine the type of vacuum corresponding to this set of critical points, we substitute the condition for a critical 
point into the potential (\ref{4.13}). Since both $\cW$ and $\partial_\Lambda \cW$  vanish at $v^2 = u_2 = 0$,  we find
\be\label{4.18}
\pV(\phi, q) \left|_{\cM_C} \right. = 0 \, .
\ee
Hence the manifold $\cM_C$ corresponds to a set of Minkowski vacua with 
vanishing cosmological constant. 

We now calculate the masses of the scalars in our model. These are given by the eigenvalues of the mass matrix
\be\label{4.19}
\cM^\Lambda_{~~\Sigma} = \left. g^{\Lambda \Xi} \, \frac{\partial}{\partial \phi^\Xi} \, \frac{\partial}{\partial 
\phi^\Sigma} \, {\rm g}^2 \, \pV(\phi, q) \right|_{\cM_C} \, ,
\ee
where $g^{\Lambda \Sigma}$ is given in eq. (\ref{4.14}). Evaluating this expression for the potential (\ref{4.13}) we find 
\be\label{4.21}
\cM^\Lambda_{~~\Sigma} = 
 (m_{\rm t})^2 \, {\rm diag} \left[ 
0 \, ,\, 1 \, , \, 0 \, , \,  1 \, , \,  0 \, , \, 1 \, , \, 0 \, , \, 1 \, , \, 0 \, , \, 0 \, 
\right], \, 
\ee
given with respect to the basis (\ref{4.15}). This result shows that the universal hypermultiplet $ v^1 , 
u_1 $ and the vector multiplet scalars $U, W$ are massless and parametrize the flat directions of the potential. 
The masses of the transition states $ v^2 , u_2  $ are given by
\be\label{4.22}
 (m_{\rm t})^2 = \frac{3}{2} \, {\rm g}^2 \,  6^{-2/3} ( U - W)^2 =
\frac{3}{2} \, {\rm g}^2 \, (h^1)^2 \, .
\ee
In terms of the microscopic picture $|h^1| = 6^{-1/3} \, |U-W|$ corresponds to the volume of the
shrinking cycle. This implies (\ref{4.22}) has precisely the structure 
expected from the eleven-dimensional point of view. By comparing
with (\ref{BPS_mass}) and using (\ref{Defv}) together with
the value $T_{(2)} = ( \ft{8 \pi}{\kappa_{(11)}^2})^{1/3}$ 
of the M2-brane tension \cite{flop}, we find that ${\rm g}$ is
fixed by M-theory,
\be
{\rm g} = \sqrt{\ft23} (48 \, \pi)^{1/3} \;.
\label{4.23}
\ee
Thus the In-picture LEEA is completely fixed once we
choose the hypermultiplet manifold to be $X(2)$.

\end{subsection}
\begin{subsection}{The scalar potential}
One of the important features of the In-picture Lagrangian  is that including the transition states gives 
rise to a scalar potential. We found that the critical points of this potential parametrize a 
submanifold $\cM_C$ which is characterized by vanishing transition states. At these points the potential vanishes 
identically.
It is then natural to ask about the properties of the potential for non-zero transition states. Especially its behavior at 
the boundaries of the  scalar manifolds is of particular interest and will be investigated in this subsection.

We start by studying the potential in terms of the vector multiplet scalars $U$ and $W$, freezing the 
hypermultiplet scalars at a fixed, non-zero value.  
\begin{figure}[t]
\renewcommand{\baselinestretch}{1}
\begin{center}
\leavevmode
\epsfxsize=0.45\textwidth
\epsffile{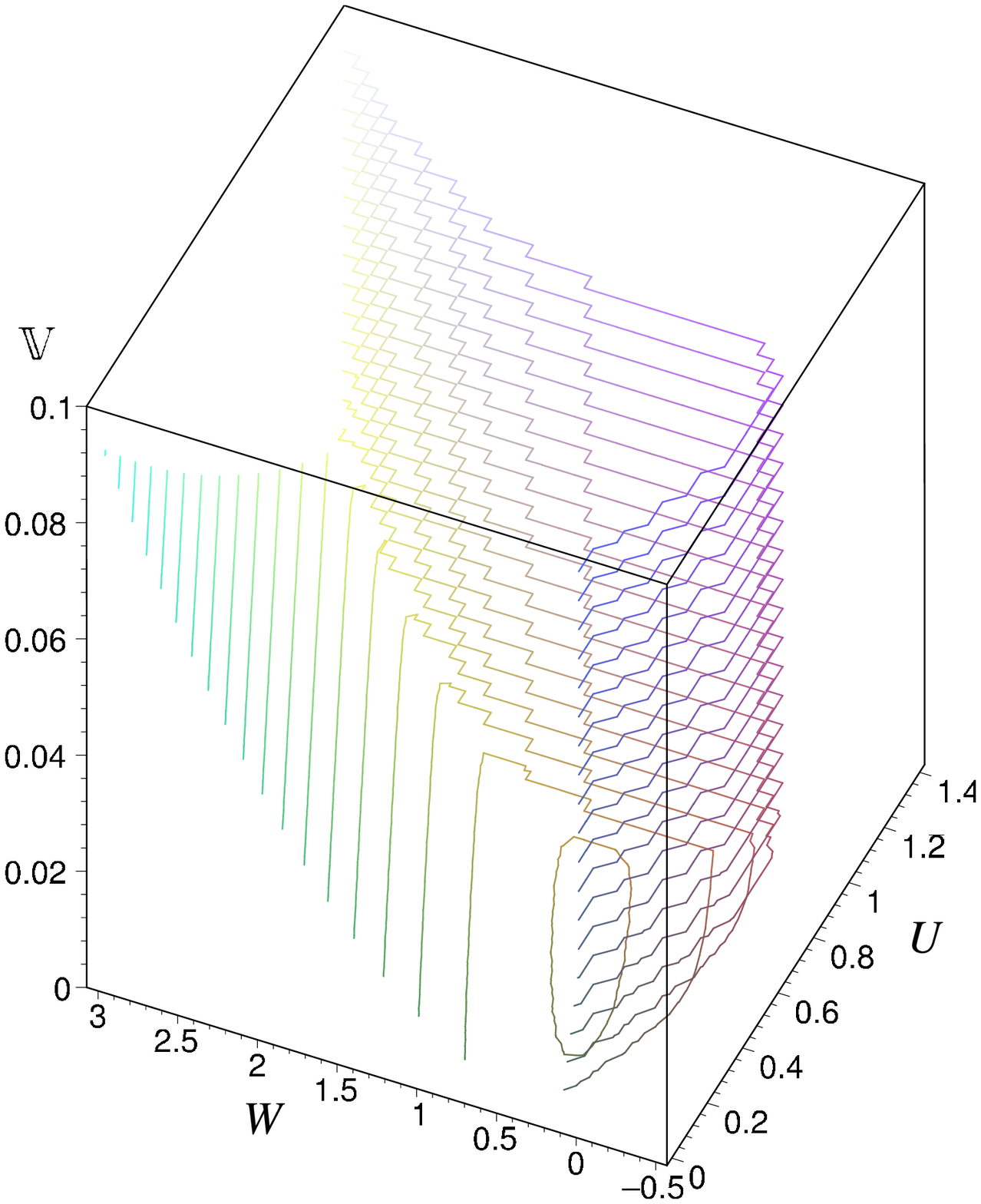} \, \, \, \, \,
\epsfxsize=0.45\textwidth
\epsffile{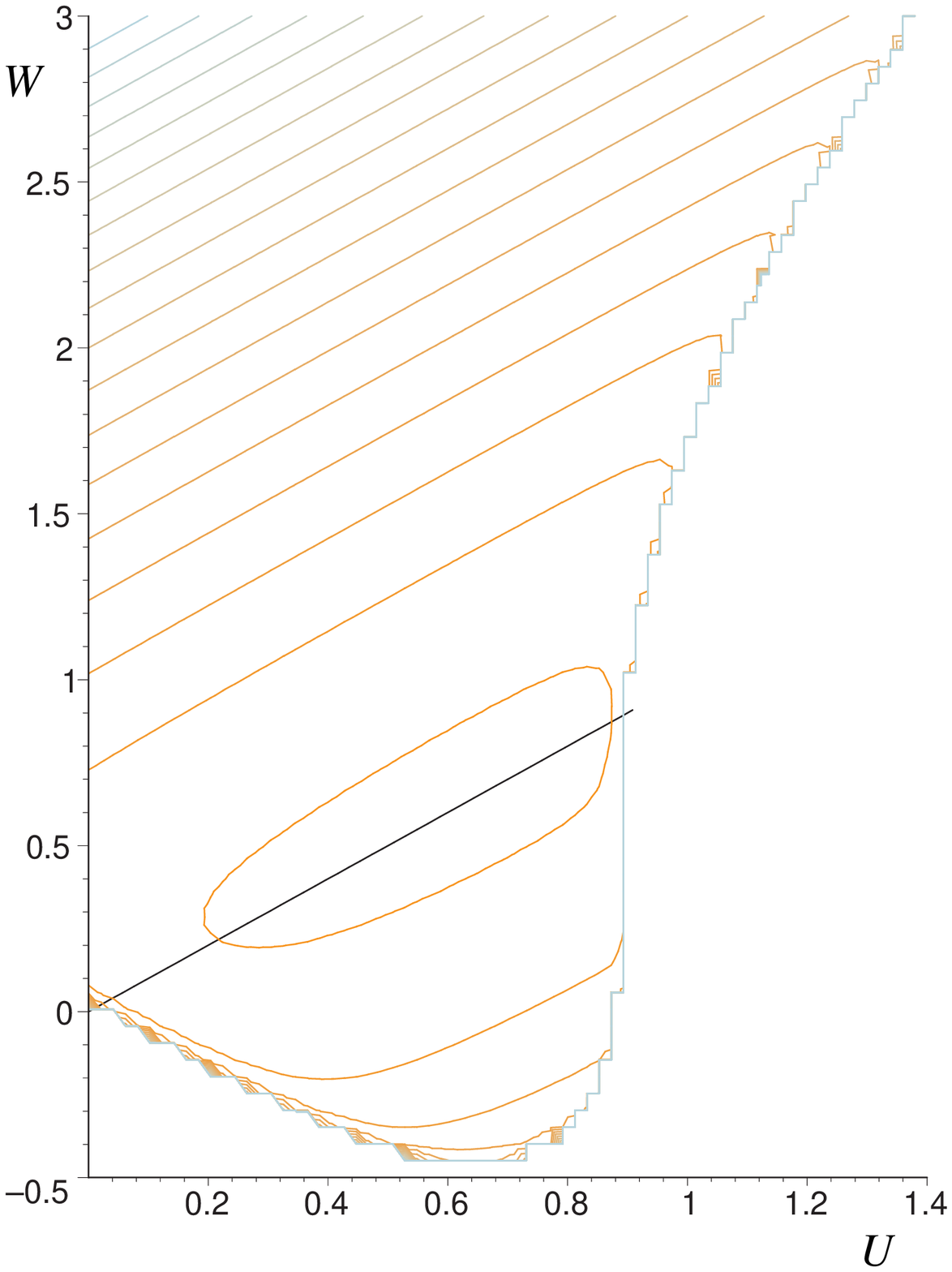}
\end{center}
\parbox[c]{\textwidth}{\caption{\label{zwei}{\footnotesize The potential $\pV(\phi, q)$ at fixed values of the 
hypermultiplet scalars $u_1 = u_2 = 0.2, v^1 = v^2 = 0$. }}}
\end{figure}
As Fig. \ref{zwei} shows, the potential is positive definite and finite as long as we are inside the vector multiplet scalar 
manifold illustrated in Fig. \ref{eins}. The potential diverges at the boundary $b_2$ where the vector multiplet metric 
$g_{xy}$ has a zero eigenvalue. At the boundary $b_1$, where $\det(g_{xy})$ is  infinite, the potential is finite. This 
behavior can be traced back to the second term of the scalar potential (\ref{2.20}) which contains the inverse metric 
$g^{xy}$.  

There are, however, additional features of the potential, which cannot be inferred from Fig. \ref{zwei} directly. While Fig. 
\ref{zwei} clearly shows that the value of the potential is small in the vicinity of the flop line $U=W$, an explicit 
calculation reveals that its actual minimum (for these fixed values of the hypermultiplet scalars) is {\it not located at} the flop line but slightly next to it. 
One should also note that even though Fig. \ref{zwei} suggests that this point is a critical point, this is not the case, 
since the derivatives of $\pV$ with respect to the hypermultiplet scalars do not vanish. Finally we observe that the 
potential diverges quadratically, $\pV \propto W^2$, in the limit $W \rightarrow \infty$. 

After analyzing the behavior of the potential at the boundaries of the vector multiplet scalar manifold, we now turn to the 
boundaries appearing in the hypermultiplet sector. These are given by the loci where the hypermultiplet metric (\ref{3.30}) 
has an infinite eigenvalue, due to  $\phi_+$ or $\phi_-$ defined in (\ref{3.25}) becoming zero. Assuming $v^1 = u_1 = 0$ and $v^2 = \vb^2 = p, u_2 = \ub_2 = q$ to be real we obtain
\be\label{4.28}
\phi_+ = 1 - p^2 \, , \quad \phi_- = 1 - q^2 + p^2 \, q^2 \, .
\ee
This shows that $p$ is bounded and takes values $-1 < p < 1$ while $q$ is unbounded. 
The dependence of the potential $\pV$ on the hypermultiplet scalars $p,q$ for frozen vector multiplet scalar fields is shown in Fig. \ref{drei}.
\begin{figure}[t]
\renewcommand{\baselinestretch}{1}
\begin{center}
\leavevmode
\epsfxsize=0.45\textwidth
\epsffile{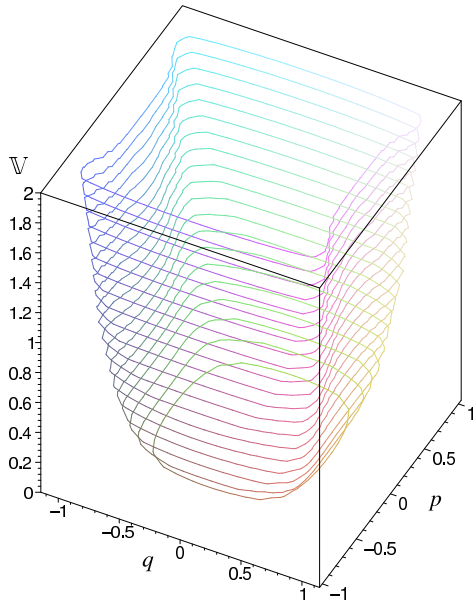}
\end{center}
\parbox[c]{\textwidth}{\caption{\label{drei}{\footnotesize The scalar potential restricted to the submanifold $U=W=0.6$, $v^1 = u_1 = 0$, $v^2 = \vb^2 = p$ and $u_2 = \ub_2 = q$. 
The potential diverges at the boundaries where $\phi_+$ or $\phi_-$ become zero.}}}
\end{figure}
This figure illustrates that $\pV$ diverges at the boundaries of the hypermultiplet moduli space where $\phi_+$ or $\phi_-$ 
vanish.  The minimum of the potential, $\pV_{\rm 
min} = 0$, is located at $q = p = 0$, which corresponds to the case of vanishing transition states.

Combined with the result obtained for the vector multiplet scalar manifold, this shows that the potential diverges at  
all boundaries of the moduli space where the scalar metrics develop a zero eigenvalue. At the boundary $U = 0$, where  $\det(g_{xy})$ is 
infinite the potential is finite.
\end{subsection}
%
%
\end{section}
\begin{section}{General flop transitions}
In the previous section we constructed the In-picture Lagrangian for a particular example of a flop transition where the 
transition states were given by a single charged hypermultiplet. We will now generalize this construction to a generic flop 
where $N$ hypermultiplets become massless at the transition locus. 
Our construction does not depend on the details of the vector multiplet scalar manifolds connected by the flop and can easily be adjusted to any specific transition. The
relation between Out-picture and In-picture is given by the orbit
sum rule (\ref{OSR}).
After fixing the hypermultiplet scalar manifold to be 
$X(N+1)$, we find that the resulting hypermultiplet sector of these In-picture Lagrangians is still uniquely determined by the 
microscopic theory.
\begin{subsection}{Constructing the action}
We will now construct the In-picture Lagrangians for a generic flop transition where the homology class $C^\star = q_I C^I$ contains 
$N$ isolated holomorphic curves. In this case the transition states are given by $N$ 
hypermultiplets, which are charged with respect to the vector field
$A^\star_\mu = q_I A^I_\mu$ associated to $C^\star$.\footnote{We do not
assume that an adapted parametrization  of the K\"ahler cone
(where $h^I > 0$) has been chosen.}
 Generalizing our 
construction from the previous section, we take the hypermultiplet scalar manifold to be $X(N+1)$, which will contain the 
universal hypermultiplet $ v^1, u_1 $ and $N$ charged hypermultiplets $ v^\alpha, u_\alpha $.  Here the index 
$\alpha = 2, \ldots, N+1$ enumerates the charged hypermultiplets which correspond to the transition states. We further use 
$h^\star = q_I h^I$ to denote the volume of the shrinking cycle $C^\star$. 

The microscopic theory imposes that the transition states are charged with respect to $A^\star_\mu$, 
while the universal hypermultiplet remains neutral. This condition requires the existence of a holomorphic  Killing 
vector of the form: 
\be\label{6.1}
k^{v^i}_{\rm gauge}  = - \imag \left[ \, 0 \, , \, v^2 \, , \,  \ldots \, , v^{N+1} \,   \right]^{\rm T} \, , \quad 
k^{u_i}_{\rm gauge}  = \imag \left[ \, 0 \, , \, u_2 \, , \,  \ldots \, , u_{N+1} \,   \right]^{\rm T} \, .
\ee
Since this Killing vector is holomorphic, its components $k^{\vb^i}_{\rm gauge}$ and $k^{\ub_i}_{\rm gauge}$ can be obtained 
from $k^{v^i}_{\rm gauge}$ and $k^{u_i}_{\rm gauge}$  by complex conjugation.  The sign conventions and overall scale in 
(\ref{6.1}) are chosen such that for $N = 1$ we reproduce the results of the previous section.

The first step is to check whether there exists a generator $t$ which gives rise to this Killing vector.  By inspection of (\ref{3.42}) we find this generator should correspond to an element of the Cartan subgroup of 
$SU(N+1, 2)$, the isometry group of $X(N+1)$. This implies that $t$ should be diagonal. In this case the general expression 
for a Killing vector on $X(N+1)$ (\ref{3.42}) simplifies to
\be\label{6.2}
\begin{split}
k^\zeta & = 0 \, , \\
k_z & = \frac{\imag}{2} \left( t^{n+1}_{~~~n+1} - t^n_{~~n}\right) \, , \\
\hat{k}^{v^i} & = \imag \left( t^i_{~j} v^j - v^i \, t^{n+1}_{~~~n+1} \right) \, ,  \\
\hat{k}^{u_i} & = \imag \left( t^{n+1}_{~~~n+1} u_i - t^j_{~i} u_j - u_i \left( t^{n+1}_{~~~n+1} - t^{n}_{~~n} \right) 
\right) \, .
\end{split}
\ee
Here it is implicitly understood that $t^i_{~j} = a_k \, \delta^i_{~j}$ is diagonal. The $a_k$ are $N+1$ real constants. 
Comparing coefficients between the  eqs. (\ref{6.1}) and (\ref{6.2}), we find the following relations for the entries of $t$:
\be\label{6.3}
\begin{split}
a_1 = t^{n+1}_{~~~n+1} \quad ,  \quad -a_\alpha + t^{n+1}_{~~~n+1} = 1 \, &  , \quad
t^{n+1}_{~~~n+1} =   t^{n}_{~n} \, , \\
 a_1 + \sum_{\alpha=2}^{N+1} a_\alpha + t^n_{~n} + t^{n+1}_{~~~n+1} =   0 \, . & 
\end{split}
\ee
The last equation arises from the condition that $t$ should be traceless. 
This set of equations has the unique solution:
\be\label{6.4}
t^1_{~1} =  t^n_{~n} = t^{n+1}_{~~~n+1} = \frac{N}{N+3} \, , \quad t^2_{~2} = \ldots = t^{N+1}_{~~~~N+1} = - \,  
\frac{3}{N+3}  \, .
\ee
Hence the gauge generator $t_{\rm gauge}$ is uniquely determined, 
\be\label{6.5}
t_{\rm gauge}  = {\rm diag} \left[ 
\frac{N}{N + 3} \, , \underbrace{ - \, \frac{3}{N+3} \, , \, \ldots \, , \, - \frac{3}{N+3}}_{N  {\rm times}} \, , \, 
\frac{N}{N + 3} \, , \, \frac{N}{N + 3} \, \right]  \, . 
\ee
Observe that in the case $N=1$, this  is exactly the generator (\ref{4.2}) of our example.

In the next step we calculate the moment map for this isometry by substituting $t_{\rm gauge}$ into eq. (\ref{3.45}). Taking 
linear combinations $\hat{\mu}^1 = \hat{\mu}^+ + \hat{\mu}^-$, $\hat{\mu}^- = - \imag \left( \hat{\mu}^+ - \hat{\mu}^- 
\right)$ and using the definition (\ref{3.25}) to simplify the resulting expressions, we obtain 
the following $SU(2)$ triplet of moment maps:
\be\label{6.6}
\hat{\mu}^r_{\rm gauge} = \left[ 
\begin{array}{c}
\frac{\imag}{2 \phi_+^{1/2} \phi_-^{1/2}} \left( v^\alpha \, u_\alpha - \vb^\alpha \ub_\alpha \right) \\
\frac{1}{2 \phi_+^{1/2} \phi_-^{1/2}} \left( v^\alpha \, u_\alpha + \vb^\alpha \ub_\alpha \right) \\
\frac{1}{2 \phi_-} \left( \ub_\alpha u_\alpha \right) - \frac{1}{2 \phi_+} \left( \vb^\alpha v^\alpha \right)
\end{array}
\right] \, .
\ee
The gauging of this isometry exactly proceeds as in the example of the previous section.

In order to complete the construction of our In-picture Lagrangians we still have to calculate the scalar potential. For 
this purpose we first derive the superpotential $\cW$ (\ref{2.21}). The moment map $P^r$ is given by 
\be\label{6.7}
P^r = \frac{1}{2} \,  h^\star \, 
\hat{\mu}^r_{\rm gauge}(q)  
\, .
\ee
Substituting in the explicit form of $\hat{\mu}^r_{\rm gauge}$, the superpotential $\cW$ becomes
\be\label{6.9}
\cW = \frac{1}{\sqrt{6}} \, h^\star \, \left( \left( \frac{1}{2 \phi_-} \left( \ub_\alpha u_\alpha\right) - \frac{1}{2 
\phi_+} (\vb^\alpha v^\alpha) \right)^2 + \frac{1}{\phi_+ \phi_-} \left( \vb^\alpha \ub_{\alpha}  \right) \left( v^\alpha 
u_\alpha \right) \right)^{1/2} \, .
\ee
Looking at $Q^r$ defined in (\ref{2.22}), we see that
\be\label{6.10}
Q^r = \sqrt{\frac{2}{3}} \, \frac{P^r}{\cW} = \frac{\hat{\mu}^r_{\rm gauge}(q)}{\left( \hat{\mu}^s_{\rm gauge}(q) 
\hat{\mu}^s_{\rm gauge}(q)  \right)^{1/2}}
\ee
is independent of the vector multiplet scalar fields and satisfies the condition $\partial_x Q^r = 0$. Hence the scalar 
potential can be expressed in terms of the superpotential and takes the form (\ref{2.23}).
\end{subsection}
\begin{subsection}{Calculating the mass matrix}
After constructing the effective Lagrangian which includes the transition states for a generic flop transition, we will now check 
that the masses of the scalar fields satisfy the conditions arising from the microscopic theory. We start by  
determining the vacuum of our theory. As in the $\mathbbm{F}_1$-model the equation
\be\label{6.11}
\partial_\Lambda \cW = 0
\ee
is solved by setting all transition states to zero.
Corollary 3 of \cite{Vicente} assures that these are all critical points of the superpotential and therefore all supersymmetric critical points of $\pV$.\footnote{The existence of further critical points of $\pV$ depends on the explicit choice of vector multiplet scalar manifold and is therefore not addressed here.} 
 The vacuum expectation values of the vector multiplet scalars and 
the universal hypermultiplet are not determined.  With this observation we find the vacuum manifold $\cM_C$ 
of our theory:
\be\label{6.12}
\cM_C = \left\{ \begin{array}{ll} 
 \, v^\alpha \, = \, u_\alpha \, = 0  \, , \quad & \alpha = 2, \ldots , N+1 \\
 \, v^1 \, , \, u_1 \, , \,  \phi^x \, ,  \quad & \mbox{not determined by eq. (\ref{6.11}).}
\end{array}
\right.
\ee
Substituting $\cM_C$ into the superpotential, we find that $\cW$ vanishes identically. Hence we have $\pV( \phi, q)|_{\cM_C} 
= 0$ and the vacuum is Minkowski. This is in complete analogy to our analysis in subsection 4.2.  

We will now calculate the mass matrix (\ref{4.19}) for our Lagrangian. In this case it is more convenient to start from the 
scalar potential in the form (\ref{2.20}):
\be\label{6.13}
\pV(\phi, q) = -4 P^r P^r + 2 g^{xy} P^r_x P^r_y  + \frac{3}{4} g_{XY} K^X K^Y \, .
\ee
Here the first observation is that for the $P^r$ given in (\ref{6.7}) the terms $P^r P^r$ and $g^{xy} P^r_x P^r_y $ are of 
fourth order in the transition states. This implies that these terms do not contribute to the mass matrix of our model since 
they vanish identically when taking two derivatives with respect to any scalar field and restricting to $\cM_C$ afterwards. 
Hence  the masses of our fields are solely generated by the last term in eq. (\ref{6.13}).

In the next step we show that the vector multiplet scalar fields $\phi^x$ are massless. The matrix
\be\label{6.14}
\cM_{\Lambda \Sigma} := \left. \partial_\Lambda \partial_\Sigma \left( 
\frac{3}{4} \, {\rm g}^2 \, g_{XY} \,  K^X \, K^Y 
\right) \right|_{\cM_C}
\ee
has non-trivial entries iff both $\Lambda$ and $\Sigma$ take values in the hypermultiplet sector. To see this, we 
expand $K^X = h^{\star}(\phi) k^X_{\rm gauge}(q)$ and note that $k^X_{\rm gauge}(q)$ vanishes when restricted to $\cM_C$. 
%
%
%
%
%
%
This implies $\cM_{\Lambda \Sigma}$ is only non-trivial if there is one derivative acting on each of the Killing vectors $k^X_{\rm gauge}(q)$.
Since $g^{\Xi \Lambda} =  g^{XY} \oplus g^{xy}$ is the 
direct sum of the hypermultiplet and vector multiplet inverse metrics, we find that  non-trivial entries of the mass matrix 
(\ref{4.19}) may occur in hypermultiplet sector only. This establishes that the vector multiplet scalars $\phi^x$ are 
massless.

Thus we restrict our analysis to the case where both $\Lambda$ and $\Sigma$ take values in the hypermultiplet sector and 
calculate the masses of the hypermultiplets. Only terms where each Killing vector is acted on by a derivative contribute to $\cM_{XY}$: 
\be\label{6.17}
\cM_{WZ} = \left. \frac{3}{2} \, {\rm g}^2 \, (h^\star)^2 \,  g_{XY} \, \partial_W \, k^X_{\rm gauge} \, \partial_Z \, k^Y_{\rm 
gauge} \, \right|_{\cM_C} \, .
\ee
The actual calculation of $\cM_{XY}$ proceeds in two steps. We first calculate the matrix $K^Y_{~X} := \partial_X k^Y_{\rm 
gauge}(q)|_{\cM_C}$. With respect to the basis
\be\label{6.18}
q^X = \left\{ \, v^1 \,, \ldots  , v^{N+1} \, , \, \vb^{1} \, , \ldots  , \, \vb^{N+1} \, , \, u_1 \, , \ldots  , \, u_{N+1} 
\, , \, \ub_1 \, , \ldots  , \, \ub_{N+1} \,  \right\} 
\ee
$K^Y_{~X}$ is diagonal and has the following form: 
\be\label{6.19}
K_Y^{~X} = {\rm diag} \left[ 0  ,  \underbrace{-\imag  , \ldots , - \imag}_{N \, {\rm times}} , 0  ,  \underbrace{\imag  , 
\ldots , \imag}_{N \, {\rm times}} , 0  ,  \underbrace{\imag  , \ldots , \imag}_{N \, {\rm times}} , 0  ,  
\underbrace{-\imag  , \ldots , -\imag}_{N \, {\rm times}}  \right] \, .
\ee

In the second step we calculate $g_{XY}(q) |_{\cM_C}$ by restricting the general expression for $g_{XY}(q)$ given in eq. 
(\ref{3.31}) to  $\cM_C$. We find that all blocks appearing in (\ref{3.30}) become diagonal. Taking into account the relation (\ref{3.50}), their non-vanishing entries are 
given by
\be\label{6.20}
\begin{array}{ll}
g_{v_1 \vb_1}  =  \frac{1}{2 \phi_+^2 \phi_-^2} \left( 1 - \ub_1 u_1  \left( 1 - \vb^1 v^1 \right)^2 \right)  \, , \quad &
 g_{v^{\alpha} \vb^{\beta}}   =   \frac{1}{2 \phi_+} \, \delta_{\alpha \beta} \, , \\
g_{u_1 \ub_1}   =  \frac{1}{2 \phi_-^2} \left( 1 - \vb^1 v^1 \right)  \,  , &
g_{u_\alpha \ub_\beta}  =   \frac{1}{2 \phi_-} \, \delta^{\alpha \beta}  \, ,
\\
g_{\vb^1 u_1}  =   - \,  \frac{1}{2 \phi_-^2} \left( \ub_1 v^1 \right) \, . & 
\end{array}
\ee
Here and in the following $\phi_+$ and $\phi_-$ are understood to be restricted to $\cM_C$. The matrix $\cM_{XY}$ can now be 
computed from 
\be\label{6.21}
\cM_{XY} = \, \frac{3}{2} {\rm g}^2 \, (h^\star)^2 \left[ K \, g \, K^{\rm T} \right] \, .
\ee
Explicitly, we find
\be\label{6.22}
\cM_{XY} =  \, 
\left[ 
\begin{array}{cccc}
 0 & A & 0 & 0 \\
 A & 0 & 0 & 0 \\
 0 & 0 & 0 & B \\
 0 & 0 & B & 0 \\
\end{array}
\right] \, ,
\ee
with $A$ and $B$ being the following $(N+1) \times (N+1)$-dimensional block matrices:
\be\label{6.22a}
A =  \frac{3}{4 \phi_+} {\rm g}^2 (h^\star)^2 \;   {\rm diag} \left[ 0  ,  \underbrace{ 1  , \ldots , 1}_{N \, {\rm times}}  
\right] \, ,  \quad B =   \frac{3}{4 \phi_-} {\rm g}^2 (h^\star)^2 \; {\rm diag} \left[ 0  ,  \underbrace{ 1  , \ldots , 
1}_{N \, {\rm times}}  \right] \, .
\ee

Finally we need to calculate the inverse metric $g^{XY}$, restricted to $\cM_C$, by inverting $g_{XY} |_{\cM_C}$ given in 
(\ref{6.20}). The resulting inverse metric is again of the structure (\ref{3.30}) with block diagonal entries. The only 
non-zero components are given by
\be\label{6.24}
\begin{array}{ll} 
  g^{v^1 \vb^1}    =   2 \, \phi_+^2 \, \phi_-  \, , \quad &
 g^{v^\alpha \vb^\beta}   =   2 \phi_+ \delta^{\alpha \beta} \, , \\
 g^{u_1 \ub_1 }  =  2 \, \frac{\phi_-}{\phi_+} \left( 1 - \ub_1 u_1 \left( 1 - \vb^1 v^1 \right)^2 \right) \, , \quad & 
g^{u_\alpha \ub_\beta}   =  2 \, \phi_- \, \delta^{\alpha \beta} \, , \\
g^{v^1 \ub_1}   =  2 \phi_+ \, \phi_- \left( \ub_1 v^1 \right) \, .  &
\end{array}
\ee

The hypermultiplet masses are given by the eigenvalues of the mass matrix
\be\label{6.25}
\cM^X_{~~Y} = \left.g^{XZ} \, \cM_{ZY} \right|_{\cM_C} \, .
\ee
Using the results (\ref{6.22}) and (\ref{6.24}), we find that the resulting matrix is 
diagonal
\be\label{6.26}
 \cM^X_{~Y} =  (m_t)^2 \, 
\left[ 
 0  ,  \underbrace{1  , \ldots , 1}_{N \, {\rm times}} , 0  ,  \underbrace{1  , \ldots , 1}_{N \, {\rm times}} , 0  ,  
\underbrace{1  , \ldots , 1}_{N \, {\rm times}} , 0  ,  \underbrace{1  , \ldots , 1}_{N \, {\rm times}}
\right]^X_{~Y} \, ,
\ee
where $(m_t)^2 = \frac{3}{2} \, (h^\star)^2 \, {\rm g}^2$.  This result explicitly shows that our Lagrangian contains one massless 
hypermultiplet, given by the complex fields $ v^1, u_1 $. This multiplet corresponds to the universal hypermultiplet. 
The transition states $ v^\alpha, u_{\alpha} $ all acquire the same mass
\be\label{6.27}
m_t = \sqrt{\frac{3}{2}} \, {\rm g} \, h^\star  \, .
\ee
It is proportional to the volume of the flopped cycle, $h^\star$, as required by the underlying 
microscopic theory. Comparing (\ref{6.27}) to eq. (\ref{BPS_mass}) we find that the gauge coupling constant ${\rm g}$ is again set by (\ref{4.23}). 
%
%

This result concludes the construction of the In-picture Lagrangian for a generic flop transition. We find that after 
fixing the hypermultiplet scalar manifold to be $X(N+1)$, the hypermultiplet sector of the resulting action is uniquely 
determined in terms of the microscopic theory. We further note that in order to calculate the mass matrix, we did not need 
to specify the details of the vector multiplet sector. Hence the analysis in this section can be used to model any  
flop transition where $N$ charged hypermultiplets become massless. In 
the case where $N = 1$ these results exactly match the ones found in the explicit example given in section 4.
\end{subsection}
\end{section}
\begin{section}{Discussion and Outlook}
In this paper we have constructed a family of 
five-dimensional gauged supergravity actions which 
can be used to describe flop transitions in M-theory
compactifications on Calabi-Yau threefolds. The new feature of these actions is that they explicitly include the extra light modes occurring in the transition region. The masses of these modes are encoded in the scalar potential. While
the vector multiplet sector could be treated exactly,
we used a toy model based on the Wolf spaces 
$X(1+N) = \frac{U(1+N,2)}{U(1+N) \times U(2)}$ to describe the hypermultiplets. In this context we worked out the metrics, the Killing vectors, and the moment maps using the superconformal quotient construction \cite{SCQ,SCgauging,SCQ2}. This geometrical data suffices to determine any hypermultiplet sector based on $X(1+N)$ in $\cN = 2$ supergravity in dimensions $d \le 6$. Furthermore, this approach considerably simplifies the investigation of gaugings, as the Killing vectors and moment maps are directly given in terms of the generators of the isometry group of the underlying Wolf space.

Our low energy effective actions have all the properties required to model a flop transition. Only the transition states acquire
a mass away from the flop and the potential has a family of degenerate
supersymmetric Minkowski ground states, which is parametrized
by the moduli of $X$.  
Therefore none of the flat directions is lifted,
and there are no additional flat directions corresponding to
Higgs branches.  Note that this is
not implied by the charge assignment alone.
The scalar potential which encodes the masses of the scalar fields is a complicated function determined by the gauging. Here it was not obvious a priori that there exists a gauging which does not lift some of the flat directions or create new ones. The latter effect could arise through
hypermultiplets combining with vector multiplets into 
long vector multiplets, giving rise to a Higgs branch.
Thus it is non-trivial that we can model a flop transition
with our quaternion-K\"ahler manifolds.

  However, it is clear that a LEEA based on $X(1+N)$ can only be a toy model, as the hypermultiplet manifolds which actually occur in string and M-theory compactifications are unlikely to be symmetric spaces.
Moreover, it is conceivable that integrating out the charged
hypermultiplets modifies the couplings of the neutral 
hypermultiplets, so that the manifolds of the In-picture and
 the Out-picture are not related by the simple
truncation \mbox{$X(1+N) \rightarrow X(1)$}.
Yet, the very fact that we find a consistent description
of a flop transition shows that while such threshold corrections
might modify the couplings, they cannot play an essential
role. This is different in the vector multiplet sector, 
where the threshold corrections play an essential role in
determining the In-picture LEEA, because the Out-picture LEEA are discontinuous.\footnote{
In the related case of $SU(2)$ enhancement it was proven in \cite{SU2} that the Out-picture LEEA cannot be extended to an $SU(2)$ invariant action without taking into account the threshold corrections.}

In summary, 
our model is a reasonable
approximation of M-theory physics 
because it (i) defines a consistent gauged
supergravity action, (ii) has, for arbitrary $N$, 
the correct properties to model a flop, (iii) is
unique (once the hypermultiplet metric is fixed) and 
(iv)  is simple enough to allow for explicit 
calculations. 
The last point will be illustrated
in a separate paper \cite{us}, where we consider cosmological
solutions.

One 
interesting direction of future research would be to take a complementary approach  and ask for the constraints imposed on a general
quaternion-K\"ahler manifold by the existence of a flop
transition. This could also be helpful for deriving such metrics from M-theory calculations. Here we
expect that again the description of quaternion-K\"ahler
manifolds in terms of hyper-K\"ahler cones is useful. 
%

Concerning the project \cite{SU2,LMZ} of deriving LEEA for topological
phase transitions and other situations with additional light
states, the next step would be to consider conifold singularities 
in type II compactifications on Calabi-Yau threefolds. As this
also involves additional massless hypermultiplets, we can use
the same hypermultiplet sector as in this paper.
The only complication is that the vector multiplet sector
 is much more involved, as it is encoded in a holomorphic
instead of a cubic prepotential. Nevertheless, we expect that the
threshold corrections can be treated along the lines of
\cite{LMZ}. Further steps would be to consider phase transitions
which have additional flat directions, such as conifold 
transitions and extremal transitions, and to include fluxes. 
The last point is of particular interest, since gaugings induced by flux are complementary to
those related to transition states, as they  
involve non-compact isometries.

Ultimately, we need to know what are the most general
gauged supergravity actions that can be obtained by
the compactification of string or M-theory including
all kinds of fluxes, branes and topological transitions.
Only once this point has been mastered, we will have
the technical tools to fully access the dynamics of 
transition states and to study their impact on problems
such as moduli stabilization, inflation, and the naturalness
problems associated with the electroweak scale and the
cosmological constant. 

\end{section}
\subsection*{Acknowledgments}
We would like to  thank
B. de Wit,
S. Vandoren
and A. Van Proeyen
for useful discussions. This work is supported by
the DFG within the `Schwerpunktprogramm Stringtheorie'. 
F.S. acknowledges a scholarship from the 
`Studienstiftung des deutschen Volkes'. L.J. was also supported by the Estonian Science Foundation Grant No 5026.

\end{document}